\documentclass[12pt]{article}

\usepackage{amssymb,amsmath} 
\pdfoutput=1
\usepackage{graphicx}
\usepackage{epsfig} 
\usepackage{latexsym}
\usepackage{psfrag}   
\usepackage{subfigure}  
\usepackage{booktabs}  
\usepackage{braket}
\usepackage{mathtools}
\usepackage{protosem}
\usepackage{wasysym}
\usepackage[numbers,sort&compress]{natbib}
\usepackage{textcomp}
\usepackage[utf8]{inputenc}
\usepackage{bbm} 
\usepackage[
      colorlinks=false,
      linkcolor=darkblue,  
      urlcolor=blue,    
      filecolor=blue,     
      citecolor=darkgreen, 
      linktocpage=true,
      pdfstartview=FitV,
      bookmarksopen=true    
      ]{hyperref}

\usepackage{mathrsfs}
\usepackage[inner=2.9cm,outer=2cm]{geometry}
\usepackage[toc]{appendix}
\usepackage{color,soul} 
\usepackage{datetime}

\usepackage{setspace}
\definecolor{darkblue}{rgb}{0.2, 0, 0.8}
\definecolor{darkgreen}{rgb}{0.2, 0.71, 0}

\newcommand{\subf}[2]{%
  {\small\begin{tabular}[t]{@{}c@{}}
  #1\\#2
  \end{tabular}}%
}
\numberwithin{equation}{section}

\newcommand{\bea}{\begin{eqnarray}}
\newcommand{\eea}{\end{eqnarray}}
\newcommand{\ba}{\begin{eqnarray}}
\newcommand{\ea}{\end{eqnarray}}

\newcommand{\beq}{\begin{equation}}
\newcommand{\eeq}{\end{equation} }
\newcommand{\beqa}{\begin{eqnarray}}
\newcommand{\eeqa}{\end{eqnarray}}
\newcommand{\beqar}{\begin{eqnarray*}}
\newcommand{\eeqar}{\end{eqnarray*}}

\newenvironment{changemargin}[2]{%
\begin{list}{}{%
\setlength{\topsep}{0pt}%
\setlength{\leftmargin}{#1}%
\setlength{\rightmargin}{#2}%
\setlength{\listparindent}{\parindent}%
\setlength{\itemindent}{\parindent}%
\setlength{\parsep}{\parskip}%
}%
\item[]}{\end{list}}

\begin{document}  


\begin{titlepage}

 \begin{flushright}
{\tt \small{IFT-UAM/CSIC-17-079}} \\
{\tt \small{IPhT-T17/138}} \\
\end{flushright}

\vspace*{1.2cm}

\begin{center}
{\begin{spacing}{1.5}\Large {\bf A systematic construction of microstate geometries with low angular momentum} \end{spacing}}

\vspace*{1.2cm}
\renewcommand{\thefootnote}{\alph{footnote}}
{\sl\large Iosif Bena$^{\text{\quarternote}}$, Pierre Heidmann$^{\text{\quarternote}}$ and Pedro F.~Ram\'{\i}rez$^{\text{\eighthnote}}$ }\footnotetext{{iosif.bena[at]cea.fr ; pierre.heidmann[at]cea.fr; p.f.ramirez[at]csic.es}}
\bigskip

$^{\text{\quarternote}}$Institut de Physique Th\'eorique, Universit\'e Paris Saclay, CEA\\ CNRS, F-91191 Gif-sur-Yvette, France \bigskip

$^{\text{\eighthnote}}$Instituto de F\'isica Te\'orica UAM/CSIC \\
C/ Nicol\'as Cabrera, 13-15, C.U. Cantoblanco, 28049 Madrid, Spain\\

\setcounter{footnote}{0}
\renewcommand{\thefootnote}{\arabic{footnote}}

\bigskip

\bigskip

\end{center}

\vspace*{0.1cm}

\begin{abstract}  
\begin{changemargin}{-0.95cm}{-0.95cm}

We outline a systematic procedure to obtain horizonless microstate geometries that have the same charges as three-charge five-dimensional black holes with a macroscopically-large horizon area and an arbitrarily-small angular momentum. There are two routes through which such solutions can be constructed: using multi-center Gibbons-Hawking (GH) spaces or using superstratum technology. So far the only solutions corresponding to microstate geometries for black holes with no angular momentum have been obtained via superstrata \cite{Bena:2016ypk}, and multi-center Gibbons-Hawking spaces have been believed to give rise only to microstate geometries of BMPV black holes with a large angular momentum \cite{Heidmann:2017cxt}. We perform a thorough search throughout the parameter space of smooth horizonless solutions with four GH centers and find that these have an angular momentum that is generally larger than 80\% of the cosmic censorship bound. However, we find that solutions with three GH centers and one supertube (which are smooth in six-dimensional supergravity) can have an arbitrarily-low angular momentum. Our construction thus gives a recipe to build large classes of microstate geometries for zero-angular-momentum black holes without resorting to superstratum technology. 

\end{changemargin}
\end{abstract} 

\end{titlepage}

\setcounter{tocdepth}{2}
{\small
\setlength\parskip{-0.5mm} 
\noindent\rule{15.7cm}{0.4pt}
\tableofcontents
\vspace{0.6cm}
\noindent\rule{15.7cm}{0.4pt}
}

\section{Introduction}

At zero gravitational coupling, String Theory can identify and count the microstates that give rise to the Bekenstein-Hawking entropy of black holes \cite{Strominger:1996sh}. However, the description of all these microstates at finite gravitational coupling, in the regime of parameters where the classical black hole exists, and in particular whether these microstates have a horizon or are horizonless remains an open problem. The latter possibility, which was proposed by Mathur in 2003\footnote{See \cite{Mathur:2005zp,Bena:2007kg,Mathur:2008nj} for reviews}, has been reinforced by recent information-theory based fuzzball/firewall arguments\footnote{See, for instance, \cite{Mathur:2009hf, Almheiri:2012rt}.} that establish that the only way a black hole can release information without a violation of Quantum Mechanics is if there exists a structure that modifies the physics at the scale of the horizon. 

The only construction of such a structure when gravity is present has been done in the context of the ``microstate geometries programme'' that aims to construct horizonless solutions with black hole charges purely within supergravity. Since supersymmetry significantly simplifies the equations governing the solutions, most microstate geometries that have been constructed so far correspond to supersymmetric black holes \cite{Bena:2006kb,Bena:2010gg,Bena:2016ypk,Heidmann:2017cxt}\footnote{However, one can also construct microstate geometries corresponding to extremal non-supersymmetric black holes  \cite{Goldstein:2008fq,Bena:2009en,DallAgata:2010srl} and to non-supersymmetric and non-extremal black holes \cite{Bossard:2014yta,Bena:2015drs,Bena:2016dbw}.}.

We will focus on the rotating three-charge BPS black hole in five dimensions, known as the BMPV black hole, which has two equal angular momenta satisfying the cosmic censorship bound $J_1 = J_2 \leq \sqrt{Q_1 Q_2 Q_3}$. This solution can be embedded in string theory as a black hole with three M2 brane charges, corresponding to M2 branes wrapping three 2-tori inside a 6-torus. In another duality frame, the three charges correspond to D1 and D5 branes that share a common direction, and momentum, P, along this direction. In the later duality frame one of the charges gives rise to a nontrivial fibration of an internal direction over spacetime, so the black hole and microstate geometries thereof are asymptotically ${\mathbb R}^{4,1} \times S^1$ solutions of six-dimensional supergravity.

Most supersymmetric microstate solutions for this five-dimensional black hole have a hyper-K\"ahler base space, and are obtained by resolving the black hole singularity via the blow-up of topologically-nontrivial bubbles that are supported against collapse by fluxes. Seen from this perspective, the microstate geometries programme is nothing but another example  of the way in which most singularities are resolved in String Theory \cite{Klebanov:2000hb,Polchinski:2000uf,Lin:2004nb}. The resulting solutions are smooth and horizonless.

A convenient choice of four-dimensional base space is given by the Gibbons-Hawking family of spaces, whose tri-holomorphic U(1) isometry implies that all solutions are determined by harmonic functions in ${\mathbb R}^3$ \cite{Gauntlett:2002nw,Gauntlett:2004qy,Bena:2005ni}. To obtain singularity-free horizonless solutions the poles of the harmonic functions must satisfy certain relations \cite{Bena:2005va, Berglund:2005vb, Ramirez:2016tqc}, and the sizes and positions of the bubbles are also constrained by the absence of closed timelike curves via the so-called bubble equations \cite{Bena:2005va,Denef:2000nb}.

Only a few explicit examples of smooth horizonless solutions which have the same charges an angular momenta as a BMPV black hole with a macroscopically-large horizon area are known  \cite{Bena:2006kb,Bena:2010gg,Bena:2016ypk,Heidmann:2017cxt}, 
and this is because most solutions one can construct by putting fluxes on a multi-center Gibbons-Hawking base have an angular momentum larger than the black hole cosmic censorship (cc) bound. This was first discovered in \cite{Bena:2006is}, where it was pointed out that smooth multi-center BPS solutions with a GH base with a large number of centers have angular momenta at and slightly above the cosmic censorship bound. Furthermore, in \cite{Heidmann:2017cxt}, a generic recipe was given to construct solutions with four GH centers that have angular momenta below the c.c. bound; however, when the aspect ratios of the distances between the centers are of the same order, all these solutions were found to have $J$ at 99\% of the cc bound. Thus, trying to find multi-GH-center  microstate geometries with low angular momentum appears to resemble searching for a needle in a haystack.

The first obstacle is to find an appropriate class of multi-center solutions with no closed timelike curves (ctc's). Since all bubbling solutions have charges dissolved in fluxes, and since these fluxes have different signs, the most likely outcome of trying to obtain a solution by putting random values of fluxes on various cycles is a solution with regions of positive and negative charge densities. Such solutions are not supersymmetric, and imposing a supersymmetric ansatz on them gives in general a solution with ctc's. Furthermore, since the flux on every cycle interacts with the flux on every other cycle, making sure there are no regions of negative charge density is a very complicated problem, that has not been solved yet\footnote{In \cite{Avila:2017pwi} a strategy to solve this problem will be proposed.}.  To bypass this problem, one of the authors proposed a recipe to construct generic ctc-free solutions with four centers, starting from ctc-free solutions describing three supertubes in Taub-NUT, going to a scaling limit, and performing a combination of spectral flows and gauge transformations to transform the supertube centers into smooth GH centers without introducing ctc's  \cite{Bena:2008wt}. One can then use the fact that the solutions have a scaling limit to remove certain constants in the harmonic functions and obtain asymptotically-${\mathbb R}^{4,1}$ smooth horizonless solutions with four GH centers and BH charges  \cite{Heidmann:2017cxt}. 

This recipe is efficient because it is relatively easy to obtain ctc-free solutions with three supertubes of different kinds and a GH center: unlike solutions with GH centers, the charges of these solutions come from the supertubes themselves, and hence by ensuring that the supertube charges are positive one avoids ctc's. Furthermore, since any solution with four GH centers can be transformed via spectral flows  into a solution with three supertubes and a GH center, the method of \cite{Heidmann:2017cxt} is guaranteed to yield the most generic ctc-free solutions with four GH centers. Moreover, if one performs this procedure and uses only two spectral flows, one obtains the most generic ctc-free solution with three GH centers and a single supertube, which is singular in the M2-M2-M2 (five-dimensional) duality frame but is smooth in the D1-D5-P  (six-dimensional) duality frame.

 The second obstacle is to implement a filter for solutions with angular momentum at a finite fraction of the cc bound. Indeed, starting from generic three-supertube solutions will almost always produce solutions with $J$ slightly below this bound, and hunting for solutions with a parametrically-lower $J$ is challenging. To do this one has to find physical quantities which will discriminate three-supertube solutions that will produce near-maximally spinning 4-GH-center solutions from those that will produce 4-GH-center solutions with lower angular momentum. To do this, it is useful to follow the procedure of \cite{Heidmann:2017cxt} and introduce
 the so-called {\it entropy parameter}:
 \begin{equation}
    \mathcal{H} \:\equiv\: \frac{Q_1 Q_2 Q_3 - J^2}{Q_1 Q_2 Q_3}\,,
     \label{EntropyParameter}
\end{equation}
which measures how far the microstate angular momentum is below the cosmic censorship bound of the black hole with the same charges\footnote{Of course, microstate geometries have no horizons and their angular momentum can easily be above the cosmic censorship bound \cite{Bena:2006is}, so the name ``entropy parameter'' is a bit of a misnomer. We use it nonetheless because it facilitates the comparison between the microstate geometry and the corresponding black hole.}.

We overpass these two obstacles and are able to construct  the largest known classes of scaling BPS smooth horizonless four-center solutions that have the same charges as BMPV black holes with a finite $\mathcal{H}$ parameter (typically 0.4 with a maximum around 0.6)\footnote{The only other known microstate geometry with multiple GH centers and low angular momentum \cite{Bena:2006kb} has $\mathcal{H} \sim 0.28$.}. The key difference between the geometries with four GH centers we construct and those of \cite{Heidmann:2017cxt}, which have $\mathcal{H} < 10^{-2}$, is that the aspects ratios of the new geometries are parametrically larger than one. Given that the only other known method for constructing  finite-$\mathcal{H}$ multi-GH-center microstate geometries, via mergers of clusters of bubbles \cite{Bena:2006kb} also produces solutions with parametrically-large aspect ratios, this appears to be a universal feature of multi-GH-center solutions with angular momenta significantly below the cc bound. It would be very interesting to find a deeper physical reason for this. 

Our technology can also be used to produce solutions with three GH centers and a supertube, that have zero or very small angular momentum. So far, the only method to obtain such BTZ microstate geometries has been to use superstratum technology \cite{Bena:2015bea, Bena:2016ypk}, which is technically much more difficult than the construction of solutions with a GH base space. Furthermore, this technology produces asymptotically-$AdS_3 \times S^3$ geometries \cite{Bena:2016ypk}, and extending these solutions to obtain asymptotically-flat D1-D5-P microstates is quite nontrivial \cite{BGMRSTW-II}. In contrast, our technology produce very easily large classes of smooth asymptotically-flat zero-angular-momentum black-hole microstate geometries. 

The trade-off is that the CFT dual of superstratum solutions is exactly known \cite{Bena:2015bea,Bena:2016ypk,Bena:2016agb,Giusto:2015dfa} (which makes superstrata amenable to precise holographic investigations), while the CFT dual of any solution with more than two GH centers is not known. The multi-center solutions we obtain do have a scaling limit, so they have a throat that can resemble a black hole throat to arbitrary accuracy; hence one can argue that they are dual to CFT states that have long effective strings \cite{Bena:2006kb} and therefore live in the same CFT sector as the states that count the black hole entropy. However, identifying these states precisely remains a challenging open problem. 

The method we employ reveals itself as a very powerful tool to study the spectrum of four-center microstate geometries. It will be interesting to be able to perform similar studies for even more general solutions, with an arbitrary number of centers or with the inclusion of non-Abelian fields \cite{Ramirez:2016tqc}. We plan to adress these questions in future work \cite{Avila:2017pwi, macroanalysis}.

In Section 2 we summarize the structure of the two classes of four-center solutions we study: solutions with four GH centers or with three GH centers and one supertube, and we explain how to generate them using generalized spectral flows and gauge transformations on solutions with three supertubes in Taub-NUT. In Section 3 we present an exhaustive analysis of solutions with four GH centers. We show that imposing a hierarchy of scales between the inter-center distances is a necessary ingredient to construct solutions with an angular momentum significantly below the cc bound. In Section 4 we apply the same kind of analysis on solutions with three GH centers and one supertube and construct microstate geometries for black holes with arbitrarily-small angular momentum.


\section{Supertubes and microstate geometries}

\subsection{Supersymmetric solutions with a Gibbons-Hawking base}
We work in the context of five dimensional $\mathcal{N}=1$ Supergravity coupled to two vector multiplets in the STU model\footnote{Our conventions mostly coincide with those of \cite{Bena:2007kg}. See \cite{Ortin:2015hya, Freedman:2012zz} for information about the theory and the STU model.}.  This theory has been shown to be obtained from compactification of eleven dimensional Supergravity on a Calabi-Yau threefold \cite{Antoniadis:1995vz}\footnote{ Alternatively, it can be obtained from the compactification of Heterotic Supergravity on $T^5$ followed by a truncation \cite{Cano:2017qrq, Cano:2017sqy, Cano:2016rls}.}. Its supersymmetric solutions with a compact spatial isometry are completely specified in terms of a set of 8 harmonic functions in $\mathbb{R}^3$, which we take of the form

\begin{equation}
V=q_\infty+\sum_a \frac{q_a}{r_a} \, , \hspace{0.3cm}
K^I=k^I_\infty+\sum_a \frac{k^I_a}{r_a} \, , \hspace{0.3cm}
M=m_\infty+\sum_a \frac{m_a}{r_a} \, , \hspace{0.3cm}
L_I=l^I_\infty+\sum_a \frac{l^I_a}{r_a} \, ,
\label{eq:harmfunc}
\end{equation}

\noindent
where $r_a$ is the Euclidean three-dimensional distance measured from the \emph{center} with coordinates $\vec{x}_a$, and with $I=1,2,3$. It is convenient to introduce a vector with the harmonic functions $\Gamma \equiv (V,K^I, L_I, M)$, which implicitly defines a set of vectors of \emph{asymptotic constants} $\Gamma_\infty$ and \emph{charges} $\Gamma_a$,\footnote{For instance, $\Gamma_a= \left( q_a , k^1_a,k^2_a,k^3_a,l^1_a,l^2_a,l^3_a,m_a \right)$ .} such that

\begin{equation}
\Gamma =  \Gamma_\infty + \frac{\Gamma_a}{r_a} \, .
\end{equation}

In this article we are mostly interested in the spacetime metric and its properties, so we shall focus on this aspect of the solution. We refer the reader to the Appendix \ref{sec:BPSsolutions} for a description of the complete field content and the solving of the BPS equations. The five-dimensional metric is given by
\begin{equation}
ds_{5}^2 \:=\: -\left(Z_1 Z_2 Z_3 \right)^{-2/3} \, \left( dt \:+\: k \right) ^2 \:+\: \left(Z_1 Z_2 Z_3 \right)^{1/3} \,ds_4^2,
\label{5Dmetric}
\end{equation}

\noindent
where \(ds_{4}^2\) is a four-dimensional ambipolar Gibbons-Hawking space \cite{Gibbons:2013tqa, Niehoff:2016gbi}
\begin{equation}
\label{eq:GHmetric}
ds_4^2 = V^{-1} \left( d\psi+\chi \right)^2 + V \left( dx^2 + dy^2 + dz^2 \right) \, , \qquad \star_{(3)} dV=d \chi \, ,
\end{equation}

\noindent
The warp factors \(Z_I\) and the 1-form $k$ are given by
\begin{eqnarray}
\nonumber
Z_I &=& L_I \:+\: \frac{1}{2}C_{IJK} \frac{K^J K^K}{V} \, , \\ \label{Z&kexpression}
k &=& \mu \left( d\psi \!+\! \chi \right) \:+\: \omega\, , 
\end{eqnarray}

\noindent
with $ C_{IJK} \:=\: \left| \epsilon_{IJK} \right|$ and 
\begin{eqnarray}
\mu &=& \frac{1}{6} V^{-2} C_{IJK} K^I K^J K^K \:+\: \frac{1}{2} V^{-1} K^I L_I \:+\: M \, , \\ 
\star_{(3)} d \omega &=& \langle \Gamma , d \Gamma \rangle \, .
\end{eqnarray}

\noindent
In the last expression \(\langle\:,\:\rangle\) is a symplectic product of vectors $A=\left(A^0, A^I, A_I, A_0 \right)$ defined as\footnote{Another more symmetric convention where the harmonic function $M$ is twice the one we use here is also used in the literature.}
\begin{equation}
    \langle A,B \rangle \:\equiv\: A^0 B_0-A_0 B^0 +\frac{1}{2} \left( A^I B_I-A_I B^I \right) \, .
    \label{symplprod}
\end{equation} 

The charges of the harmonic functions are usually constrained by the properties of the solution that is being described. While we will review specific restrictions for supertubes and microstate geometries later, we emphasize here that all physically sensible solutions need to be free of closed timelike curves and Dirac-Misner strings. The first condition requires the positivity of the \emph{quartic invariant} $\mathcal{I}_4$ (see Appendix \ref{sec:BPSsolutions}),

\begin{equation}
\label{eq:I4}
\mathcal{I}_4 \equiv Z_1 Z_2 Z_3 V - \mu^2 V^2 > 0\, .
\end{equation}

\noindent
 while the second restricts the position of the centers \cite{Denef:2000nb},
\begin{equation}
\label{eq:bubble1}
\sum_b \frac{ \langle \Gamma_a , \Gamma_b \rangle}{r_{ab}} = \langle\Gamma_\infty , \Gamma_a \rangle \, , \qquad \text{where} \hspace{0.5cm} \Gamma \equiv \Gamma_\infty + \sum_a \frac{\Gamma_a}{r_a} \, ,
\end{equation}

\noindent
where $r_{ab}$ is the distance between the pair of centers located at $\vec{x}_a$ and $\vec{x}_b$. These are known as the \emph{bubble equations} and impose strong constraints on the space of parameters leading to physically sensible configurations. Solving those equations is usually the hardest step when building multi-center solutions.

\subsection{Symplectic transformations}
\label{sec:transformations}

Any vector of harmonic functions defines a solution, and any linear transformation, $\Gamma ' = g \Gamma$ with $g\in GL(8,\mathbb{R})$, maps a solution to another solution. A special subgroup of these transformations is $Sp(8,\mathbb{R})$, corresponding to linear transformations that preserve the symplectic product and, therefore, leave the bubble equations invariant. Among all possible $Sp(8,\mathbb{R})$ transformations, the most attractive are those that also leave the function $\mathcal{I}_4$ invariant. We are interested in two subgroups with these characteristics \cite{Crichigno:2016lac}:

\begin{itemize}
\item \textbf{Generalized spectral flows}. These transformations can be understood as simple changes of coordinates when the five dimensional solution is embedded in six dimensional Supergravity \cite{Bena:2008wt}, and correspond to a subgroup of the $E_{7(7)}$ duality transformations from the eleven dimensional perspective \cite{DallAgata:2010srl}. Generalized spectral flows are generated by three real parameters $\gamma^I$

\begin{equation}
\begin{aligned}
&M' = M, \qquad L_{I}' = L_{I} \:-\: 2\, \gamma^I M, \\
&K^{I \, '}  = K^{I} \:-\: C_{IJK}\, \gamma^J L_K \:+\:  C_{IJK}\, \gamma^J \gamma^K M, \\
&V'  = V \:+\:  \gamma^I K^I \:-\: \frac{1}{2} C_{IJK}\, \gamma^I \gamma^J L_K \:+\: \frac{1}{3} C_{IJK} \,\gamma^I \gamma^J \gamma^K M.
\label{eq:spectralflow}
\end{aligned}
\end{equation}

Even though they act non-trivially on $Z_I$ and $\mu$, one can check that $\mathcal{I}_4$ and the bubble equations remain invariant under the action of (\ref{eq:spectralflow}).

\item \textbf{Gauge transformations}. These transformations leave the physical properties of the solution unchanged and their sole effect is a gauge transformation of the vector fields. They are just a reflection of the fact that the construction of solutions in terms of 8 harmonic functions contains redundancies. There are three independent gauge transformations (one for each vector) parametrized by $g^I$, acting as

\begin{equation}
\begin{aligned}
&V' = V, \qquad K^{I\, '} = K^{I} \:+\: g^I V, \\
&L_{I} ' = L_{I} \:-\: C_{IJK}\, g^J K^K \:-\: \frac{1}{2}  C_{IJK}\, g^J g^K V, \\
&M'=  M \:-\: \frac{1}{2} g^I L_I \:+\: \frac{1}{4} C_{IJK}\, g^I g^J K^K \:+\: \frac{1}{12} C_{IJK} \,g^I g^J g^K V,
\label{eq:gaugetransformation}
\end{aligned}
\end{equation}

As shown in \cite{Heidmann:2017cxt, Bena:2008wt}, performing a generalized spectral flows of type $I$ transforms a supertube of species $I$ to a Gibbons-Hawking center. The gauge transformations enable to get rid of the constant terms appearing after spectral flows in the functions $K^I$, since these introduce singularities for five dimensional asymptotically-flat solutions. Consequently, generalized spectral flows and gauge transformations can be used to generate smooth horizonless geometries starting from a three-supertube solution in a Taub-NUT hyper-K\"ahler space. This plays a central role in our study.

\end{itemize}

There is one additional subgroup of $Sp(8,\mathbb{R})$ that leaves $\mathcal{I}_4$ invariant that involves rescalings of the harmonic functions, but since we are not going to make use of this type of transformations we refer the interested reader to \cite{Crichigno:2016lac}.

\subsection{Three-supertube scaling BPS solutions in Taub-NUT}
\label{sec:supertubes}

Our starting point is a system of three two-charge supertubes of different species in which the $4$-dimensional hyper-K\"ahler metric is the Euclidean Taub-NUT solution \cite{Sorkin,Gross-Perry}. This is a multi-supertube generalization \cite{Vasilakis:2011ki} of the configuration constructed  in \cite{Bena:2005ay}. However, since the supertubes are of different kinds, this configuration is not smooth in the D1-D5-P duality frame.

Each supertube carries a dipole charge \(k_I\) and two electric charges \(Q_a^{(I)}\) at the centers $a\neq I$. Consequently, the $8$ harmonic functions that characterize such a field configuration are given by
\begin{eqnarray}
V&=&q_\infty+\frac{q_0}{r_0} \, , \\
K^{I} &=& \alpha^{I}+\sum_{a=1}^{3}\frac{k_a}{r_{a}}  \delta^{I}_{a}\, , \\
L_{I} &=& 1+\sum_{a=1}^{3}\frac{Q^{(I)}_a}{4r_a}\left(1-\delta^{I}_{a}\right) \, , \\
M &=& m_{\infty}+\sum_{a=0}^{3}\frac{m_a}{r_a} \, .
\end{eqnarray}

\noindent
In these expressions $r_a$ is the three-dimensional Euclidean distance measured from the a$^{th}$ center. We consider axisymmetric supertube configurations. The positions of the supertube centers are given by the distances \(z_1\), \(z_2\) and \(z_3\) on the z-axis of the three-dimensional base space of the solution in the following order
\begin{equation}
z_1 \:>\: z_2 \:>\: z_3 \:>\: z_0 \:=\: 0.
\end{equation}
Therefore,
\begin{equation}
r_a \:\equiv\: \sqrt{x^2 \!+\! y^2 \!+\! \left(z-z_a\right)^2} \, ,
\end{equation}

In the analysis performed in \cite{Bena:2009en, Vasilakis:2011ki, Heidmann:2017cxt} it was derived that regularity at the centers and the absence of asymptotic Dirac-Misner strings requires fixing the following parameters

\begin{equation}
\begin{split}
& m_1 \:=\: \frac{Q_1^{(2)}Q_1^{(3)}}{32 k^{(1)}_1},\qquad m_2 \:=\: \frac{Q_2^{(1)}Q_2^{(3)}}{32 k^{(2)}_2}, \qquad m_3 \:=\: \frac{Q_3^{(2)}Q_3^{(1)}}{32 k^{(3)}_3}, \\
& m_0 \:=\: 0, \qquad m_\infty \:=\: - \sum\limits_{a=1}^3 \frac{m_a}{z_a}, \qquad \alpha^1 \:=\: -2 q_\infty m_\infty \, , \qquad \alpha^2 \:=\: \alpha^3 \:=\: 0  \, .
\end{split}
\label{MConstants}
\end{equation}

This set of harmonic functions produces a physically sensible configuration when there are no Dirac-Misner strings between centers and no ctc's in the spacetime metric. This is achieved imposing the four bubble equations (\ref{eq:bubble1}), which fix the positions of the centers, and the global bound (\ref{eq:I4}). The bubble equations can be conveniently written as
\begin{equation}
\begin{split}
& \frac{\Gamma_{12}}{r_{12}} \:+\: \frac{\Gamma_{13}}{r_{13}} \:-\: 8 q_0 \frac{m_1}{z_1}\:=\: 8 m_1  q_\infty \:-\: 4 k_1 \, , \\
& \frac{\Gamma_{21}}{r_{12}} \:+\: \frac{\Gamma_{23}}{r_{23}} \:-\: 8 q_0 \frac{m_2}{z_2}\:=\: 8 m_2 q_\infty \:-\: 4 k_2 \:-\: 2 Q_2^{(1)} q_\infty m_\infty \, , \\
& \frac{\Gamma_{32}}{r_{23}} \:+\: \frac{\Gamma_{31}}{r_{13}} \:-\: 8 q_0 \frac{m_3}{z_3} \:=\: 8 m_3  q_\infty \:-\: 4 k_3 \:-\:2 Q_3^{(1)} q_\infty m_\infty\, , \\
\end{split}
\label{BubbleEquationSup}
\end{equation} 
where $\Gamma_{ab}=\langle \Gamma_a , \Gamma_b \rangle $ and $r_{ab}$ is the distance between the centers $a$ and $b$. Provided those conditions are satisfied, we have a family of regular solutions free of ctc's labeled by eight parameters; $k^I_a$, $Q^I_a$, $q_\infty$ and $q_0$. However, one should not expect the whole space of parameters to be compatible with (\ref{eq:I4}) and (\ref{BubbleEquationSup}). Moreover, experience shows that finding a set of appropriate parameters can involve a vast exploration.

Among all possible physical solutions, the most interesting correspond to \emph{scaling} geometries. These are configurations in which the distances between the supertubes and the GH center can be made arbitrarily small while preserving the value of the asymptotic charges practically constant. If one defines the aspect ratios $d_I$ as $z_I = \lambda d_I$ with $d_3$ of order one, this is achieved in practice for configurations in which the terms on the left-hand side of \eqref{BubbleEquationSup} are almost vanishing when we replace the inter-center distances by the aspect ratios. Thus, scaling solutions of three supertubes and a GH center must satisfy the \emph{scaling conditions}:
\begin{equation}
\begin{split}
& \frac{\Gamma_{12}}{d_{12}} \:+\: \frac{\Gamma_{13}}{d_{13}} \:-\: 8 q_0 \frac{m_1}{d_1}\:\approx\: 0\\
& \frac{\Gamma_{21}}{d_{12}} \:+\: \frac{\Gamma_{23}}{d_{23}} \:-\: 8 q_0 \frac{m_2}{d_2}\:\approx\: 0 \\
& \frac{\Gamma_{32}}{d_{23}} \:+\: \frac{\Gamma_{31}}{d_{13}} \:-\: 8 q_0 \frac{m_3}{d_3}\:\approx\: 0,\\
\end{split}
\label{ScalingConditionSup}
\end{equation}

\noindent
with $z_{IJ} = \lambda \, d_{IJ}$. When these relations are satisfied, the limit $\lambda \ll 1$ in the bubble equations is well-defined at first order in $\lambda$. By summing the three equations \eqref{ScalingConditionSup} we see that $m_1$, $m_2$ and $m_3$ cannot be all of the same sign. Since all $Q_J^{(I)}$ are taken positive to avoid ctc's one of the dipole charges $k_1$, $k_2$ and $k_3$ must have different sign from the other two. The warp factors of the solution \eqref{Z&kexpression} have a term quadratic in the dipole charges, and when the $k_I$ have opposite signs this term is negative and can be problematic. However we avoid this by choosing the $k_I$ to be smaller than the square roots of the charges.

As one approaches the scaling limit, the $AdS_2 \times S^3$ throat of the solution becomes longer, and the solution resembles more and more the near-horizon geometry of an extremal black hole. Therefore after the application of spectral flow transformations that render the metric smooth at the centers we will construct a completely regular, horizonless solution with near-horizon-like throat of large but finite depth that caps off smoothly.

\subsection{Microstate geometries from three-supertube configurations}
\label{subsec:microgeom}

Several classes of smooth BPS solutions can be generated through the application of two or three generalized spectral flows and gauge transformations on a system of three-supertubes in a Taub-NUT space, see for instance \cite{Bena:2008wt, Vasilakis:2011ki, Heidmann:2017cxt}. In this manner we can  investigate large classes of regular supersymmetric solutions with multiple Gibbons-Hawking centers, which are usually difficult to generate otherwise. As we explained in section \ref{sec:transformations}, no closed timelike curves or Dirac-Misner strings are generated in this process. 

However, this method to generate smooth microstate geometries presents a drawback. It has been recently argued that generalized spectral flows result in a significant increase of the angular momentum, at least when the inter-center distances are of the same order of magnitude for four-center solutions \cite{Heidmann:2017cxt}. Actually, when all the distances between the centers are of the same order the solutions are near-maximally spinning. In particular, while spectral flows do not modify the quartic invariant, it seems that they simultaneously increase the value of the two terms in its defining expression \eqref{eq:I4}. 

To be more precise, recall that we defined the \emph{entropy parameter} $\mathcal{H}$ in \eqref{EntropyParameter} as:
\begin{equation}
    \mathcal{H} \equiv \frac{Q_1 Q_2 Q_3 - J^2}{Q_1 Q_2 Q_3} \, ,\nonumber
\end{equation}

\noindent
where $Q_I$ and $J$ are the asymptotic charges and angular momentum. For classical black holes the entropy is proportional to horizon area, given by the square root of the numerator. The numerator can also be read off from the coefficient of the $1/r^4$ in the quartic invariant. If the numerator is negative, the black hole solution will be singular. Thus, $\mathcal{H}$ is 0 when rotation is maximal, while it is 1 when there is no rotation at all.  

Within this construction scheme, there are two possible strategies that we are going to explore in order to avoid angular momenta near the cosmic censorship bound. The first possibility is to look for configurations in which there is a hierarchy in the distances between centers. The second option simply consists in applying two spectral flows instead of three. Both approaches involve a large exploration of the parameter space, since there is no way, in principle, to know how the input parameters should be chosen to produce a high value of the entropy parameter. On the bright side, the procedures to generate smooth solutions can be systematized, as we will briefly explain, and therefore such an exploration is feasible. 

\begin{itemize}
    \item \textbf{1. Four smooth Gibbons-Hawking centers.}
\end{itemize}
If one applies the three possible generalized spectral flow transformations to an initial solution with three supertubes in Taub-NUT, one obtains a four-GH-center configuration described by a set of harmonic functions with
\begin{equation}
\begin{split}
& l_a^I \:=\: -\frac{1}{2}C_{IJK} \frac{k_a^J k_a^K}{q_a} \, ,\qquad a \in \{0,\ldots,3\} \, , \\
& m_a \:=\: \frac{1}{12}C_{IJK} \frac{k_a^I k_a^J k_a^K}{q_a^2}  \qquad a \in \{0,\ldots,3\}\,.
\label{constraintcharges}
\end{split}
\end{equation}
\noindent
This guarantees that the resulting solution is horizonless and smooth \cite{Bena:2005va,Berglund:2005vb,Bena:2007kg}. 

\begin{itemize}
    \item \textbf{2. One supertube and three smooth Gibbons-Hawking centers.}
\end{itemize}
Let us consider the application of two types of generalized spectral flows \eqref{eq:spectralflow} to an initial system of three supertubes in Taub-NUT. For instance let us denote by the index $J$ the spectral flow transformation which is not applied, so $\gamma^J=0$. Then it is straightforward to check that the set of four-center harmonic functions obtained satisfy 
\begin{equation}
\begin{split}
& l_a^I \:=\: -\frac{1}{2}C_{IKL} \frac{k_a^K k_a^L}{q_a} \qquad a\neq J\, , \\
& m_a \:=\: \frac{1}{12}C_{IKL} \frac{k_a^I k_a^K k_a^L}{q_a^2}  \qquad a \neq J\, , \\
& k^I_J \:=\: 0 \qquad I\neq J\, , \\
& l^J_J \:=\: 0\, , \\
& m_J \:=\: \frac{1}{4} C_{JKL} \frac{l_J^K l_J^L}{k^J_J}\, ,
\label{constraintcharges2}
\end{split}
\end{equation}
\noindent
where the notation is that of \eqref{eq:harmfunc}. This configuration describes a supertube in the presence of three smooth Gibbons-Hawking centers. Much like vanilla two-charge supertubes ~\cite{Balasubramanian:2000rt,Maldacena:2000dr,Mateos:2001qs, Lunin:2001fv,Emparan:2001ux}, these solutions are not smooth in the M2-M2-M2 duality frame where they are described by five-dimensional supergravity. However, they become smooth once one dualizes them to a D1-D5-P duality frame where the supertube charges correspond to D1 and D5 branes, and the solution can be described by a six-dimensional supergravity \cite{Bena:2008dw}.

\section{Four-GH-center solutions with a hierarchy of scales}

In this section we explore the possibility of constructing smooth geometries with four Gibbons-Hawking centers with an angular momentum far below the cosmic censorship bound. We will see that when the analysis technique of \cite{Heidmann:2017cxt} is applied to solutions in which the inter-center distances have a hierarchic structure, it is possible to build solutions with small angular momentum. The particular five-center solution found in \cite{Bena:2006kb} provides the motivation to study this type of configurations in detail, since it is characterized by a hierarchic distribution of the centers and its entropy parameter is $\mathcal{H}=0.28$.

\subsection{Exploration of the parameter space}

The details of the numerical analysis we perform are contained in Appendix \ref{sec:numerics}. Here we give a qualitative description of the procedure pioneered in \cite{Heidmann:2017cxt} and explain our results. The program is based on the automation of the method to build four-center microstate geometries described in the previous section. This allows us to scan the space of parameters and look for the maximization of the entropy parameter $\mathcal{H}$.

\subsubsection{Systematic generation of solutions}
\label{subsec:SystematicProcedure}

Let us discuss how the solution generating technique is automatized. Before proceeding, we point out some generalities about this construction scheme. In first place, we are interested in configurations that present a hierarchy of scales, which at the beginning we take to be
\begin{equation}
\begin{split}
\frac{z_1}{z_2} &\:\approx\: 10^2 \, , \\
\frac{z_2 }{z_3} &\:\approx\: 10^2 \, .
\end{split}
\label{distanceratios}
\end{equation}

\noindent
The Gibbons-Hawking metric \eqref{eq:GHmetric} is fully determined by the function $V$. Although it might seem that this metric becomes singular at the centers, it can be easily checked that this is not true as long as the coefficients $q_a$ are integer numbers\footnote{This fact becomes evident performing a local coordinate transformation $r_a=\frac{\rho^2_a}{4}$. The local metric describes the orbifold space $\mathbb{R}^4/\mathbb{Z}_{\vert q_a \vert}$, which is harmless in the context of string theory.}. Then, we need to take care of that fact and impose that all Gibbons-Hawking charges are integer numbers. Moreover their sum has to be necessarily 1, since we want this space to asymptote to $\mathbb{R}^4$.

On the other hand, the application of spectral flow transformations to a system of supertubes does not guarantee a good asymptotic behaviour. In particular we are interested in asymptotically-flat 5-dimensional spacetimes. However this type of configurations will never be obtained directly if we start from a system of supertubes in a Taub-NUT base space, since one cannot eliminate simultaneously all constant factors in the functions $V$ and $K^I$. The best one can do is to perform three gauge transformations \eqref{eq:gaugetransformation} to eliminate the integration constants in $K^I$, and remove by hand the constant in $V$ afterwards, hoping that this does not generate ctc's.

The initial system of supertubes is specified by seven parameters: $k_1$, $k_2$, $k_3$, $q_0$, $\frac{Q_2^{(1)}}{Q_1^{(3)}}$, $\frac{Q_3^{(2)}}{Q_1^{(2)}}$ and $\frac{Q_2^{(3)}}{Q_3^{(1)}}$. The solution also depends on $q_\infty$, but the value of this parameter is not essential when looking for scaling solutions. Our recipe is the following:

\begin{enumerate}
    \item Choose a value for the seven degrees of freedom of the three-supertube solution, $k_1$, $k_2$, $k_3$, $q_0$, $\frac{Q_2^{(1)}}{Q_1^{(3)}}$, $\frac{Q_3^{(2)}}{Q_1^{(2)}}$,$\frac{Q_2^{(3)}}{Q_3^{(1)}}$. Recall that we can only obtain scaling solutions if one of the $k$'s has a sign different from the other two. We also give a non-vanishing value to $q_\infty$, which is necessary in order to be able to cancel the constant terms of all $K^I$ in a later step. Therefore, the base space is Taub-NUT.
    \item Using \eqref{distanceratios} as an equality, we impose the scaling condition \eqref{ScalingConditionSup} as three exact equations from which we obtain the precise value of all the $Q^{(I)}_a$ parameters. Afterwards, we round these values to some close rational numbers and solve the bubble equations \eqref{BubbleEquationSup} to determine the positions of the centers $z_1$, $z_2$ and $z_3$. Thus, \eqref{distanceratios} and \eqref{ScalingConditionSup} cease to be equalities and become approximations, as intended. This step ensures that we construct a scaling three-supertube solution free of ctc's.
    \item We perform three generalized spectral flows and three gauge transformations. We fix the values of the spectral flow parameters $\gamma^I$ by imposing some particular, integer values of the Gibbons-Hawking charges $q_1$, $q_2$ and $q_3$ such that $\sum q_a=1$. The values of the gauge parameters $g_I$ are found requiring that the constant terms in all the functions $K^I$ are zero. 
    
    At this stage, we have a BPS scaling solution with four Gibbons-Hawking centers. However, there are still two problems that need to be solved. First, the harmonic function $V$ still has a constant term. This means that the four-dimensional base space of the solution is asymptotically $\mathbb{R}^3\times S^1$ instead of flat $\mathbb{R}^4$. Second, because all the parameters of the transformations $\gamma_I$ and $g_I$ are fixed by polynomial equations, the resulting charges and dipole charges of the solution are general real numbers. Since those are expected to be quantized when interpreted in the full context of string theory, it is desirable that they take integer values.
    
    For the numerical analysis of the entropy parameter, we do not apply the next three steps because they do not significantly change the value of the charges and the angular momentum. They are just technical steps to build proper asymptotically-5-dimensional solutions.
    \item It is not possible to remove the constant of $V$ using transformations that preserve the bubble equations. Thus we remove it by hand. The impact of this removal on the solution takes place mainly on the bubble equations. Changing the right hand side of the bubble equations \eqref{eq:bubble1} necessarily results in a change of the inter-center distances in the left hand side. In the scaling limit, when all these distances are very small, one may think that a change of constant terms can be compensated by an infinitesimally small change of distances. However, this it is not necessarily true for axisymmetric configurations \cite{Heidmann:2017toappear}. In our construction we will carefully select the solutions for which it is possible to perform this truncation preserving the axisymmetry of the center configuration.
    \item Since we want the monopole and dipole charges to be integer numbers, we proceed in two steps. The first step consists in obtaining solutions whose harmonic functions have rational poles. For that purpose, we round the values of the parameters $k^I_a$ to be rational and obtain all the other charges $l^I_a$ and $m_a$ using \eqref{constraintcharges}. Since one can find rational numbers arbitrarily close to any irrational number, this procedure is guaranteed not to change significantly the properties of the solution. Hence, we have a fair bit of freedom in rounding the irrational numbers to rational ones, and we can use it to obtain $k^I_a$ that have the same denominator. This rounding does not leave the bubble equations invariant, and we need to solve them again and check again the absence of ctc's. The second step is to obtain solutions whose harmonic functions have integer poles. To do this we use the following transformations parametrized by any real numbers $\{s_1,s_2,s_3\}$,
\begin{alignat}{2}
&M \:\rightarrow\: \frac{1}{6} C_{IJK} s_I s_J s_K M,\qquad &&L^{I} \:\rightarrow\: \frac{1}{2} C_{IJK} s_J s_K L^{I},\nonumber \\
&V \:\rightarrow\: V,\qquad &&K^{I} \:\rightarrow\: s_I\, K^{I}, \qquad \{s_1,s_2,s_3\}\in\mathbb{R}^3.
\label{overallmultiplication}
\end{alignat}
They preserve the regularity of the solution. Indeed, all the horizonless conditions \eqref{constraintcharges} are still satisfied and the bubble equations and the quartic invariant are multiplied by an overall factor $s_1s_2s_3$ and $(s_1s_2s_3)^2$ respectively while ${\mathcal H}$ does not change. Thus, one chooses the three $s_I$ to be the smallest integers needed to obtain integer charges from the rational charges.
    \item The factors $s_I$ are usually large numbers, so multiplying the harmonic functions $L^I$ and $M$ by them makes their constant terms very large. Asymptotic flatness of the five-dimensional metric \eqref{5Dmetric} demands having the constant terms of all $L^I$ equal to one\footnote{Actually only their product has to be equal to one, but this subtlety is not particularly relevant.}. To obtain such solutions one again has to change by hand the constant terms of all the $L^I$. As explained in \cite{Heidmann:2017toappear}, such a change can always be done for scaling solutions, and results in a global dilatation of the multicenter configuration. To make the inter-center distances small again, we simply fine-tune the value of some of the dipole charges (keeping them integer) to make the solution scale \cite{Bena:2006kb}. \end{enumerate}

This method produces asymptotically-flat, scaling solutions with four Gibbons-Hawking centers that have integer charges. Using this systematic procedure we can build a huge number of four-GH-center solutions and obtain the variation of the entropy parameter $\mathcal{H}$ as one moves in the parameter space spanned by $k_1,k_2,k_3,\frac{Q_2^{(1)}}{Q_1^{(3)}},\frac{Q_3^{(2)}}{Q_1^{(2)}},\frac{Q_2^{(3)}}{Q_3^{(1)}},q_0,q_1$ and $q_2$.

\subsubsection{Main results of the analysis}
\label{subsubsec:NumAnalysis}

We divided our analysis in three parts, considering the effect of modifying three sets of parameters: the Gibbons-Hawking charges ($q_0,q_1,q_2$), the supertube dipole charges ($k_1,k_2,k_3$) and the supertube charge ratios ($\frac{Q_2^{(1)}}{Q_1^{(3)}},\frac{Q_3^{(2)}}{Q_1^{(2)}},\frac{Q_2^{(3)}}{Q_3^{(1)}}$). We reach the following conclusions:
\begin{itemize}
\item The entropy parameter approaches zero drastically when the absolute value of the Gibbons-Hawking charges is large. The optimal value we observed for the Gibbons-Hawking charges is 1,1,1 and -2. 
\item For the initial supertube dipole charges, we observed that configurations with $k_2$ negative and $k_1$ and $k_3$ positive are the optimal ones. With the two other sign configurations, we did not find domains of charge ratios with an entropy parameter bigger than 0.1. We also noticed that the entropy parameter does not depend significantly on $k_2$ and it depends essentially on $\frac{k_1}{k_3}$. Furthermore, we observed that for any charge ratios one can find a particular dipole ratio $\frac{k_1}{k_3}$ where the entropy parameter is maximal and the upper bound seems to be $\mathcal{H}\sim 0.3$.
\item With the optimal configuration of dipole charge signs and Gibbons-Hawking charges, we have found several domains of charge ratios where the entropy parameter is above 0.2. 
\end{itemize}

\begin{figure}
\centering
\includegraphics[width=100mm]{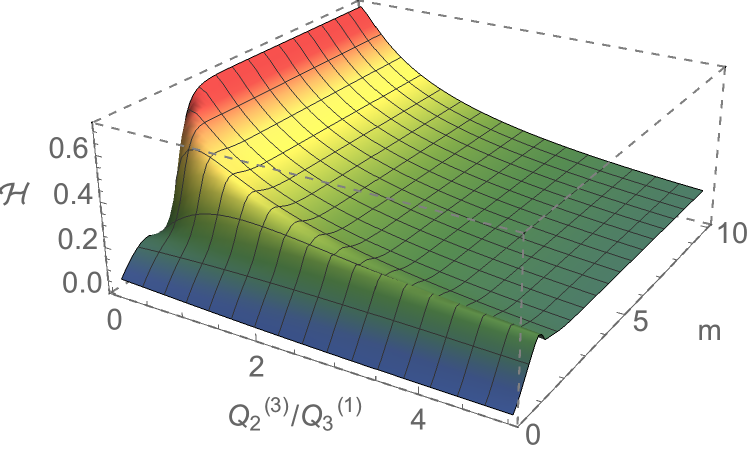}
\caption{The entropy parameter $\mathcal{H}$ as a function of $\frac{Q_2^{(3)}}{Q_3^{(1)}}$ and the initial hierarchy parameter $m=\log_{10}\left[\frac{z_1}{z_2}\right]=\log_{10}\left[\frac{z_2}{z_3}\right]$. The other parameters are fixed to the following values $q_0 = q_1=q_2= 1$, $k_1= -k_2= k_3 = 1$, $\frac{Q_2^{(1)}}{Q_1^{(3)}} = 0.9$ and $\frac{Q_3^{(2)}}{Q_1^{(2)}} = 2$.}
\label{graphs_distanceratio}
\end{figure}

Moreover, we performed an analysis to study the impact of the hierarchy of scales. In Figure \ref{graphs_distanceratio}, we show one of the main results of the analysis. It illustrates how the entropy parameter can significantly increase with the aspect ratios. The entropy parameter is represented with respect to two variables, one of the charge ratios and the order of magnitude of the hierarchy $m$, which is defined as
\begin{equation}
\begin{split}
\frac{z_1}{z_2} &= 10^m \\
\frac{z_2 }{z_3} &= 10^m.
\end{split}
\end{equation}
The rest of parameters are chosen to optimize the entropy parameter, according to the numeric results just presented (see Appendix \ref{sec:numerics} for more details). The graph shows that when $m$ is around 0 the solutions are near-maximally spinning, with $\mathcal{H}$ very close to 0, recovering the results of \cite{Heidmann:2017cxt}. Furthermore, in all the solutions we examined the entropy parameter increases as the hierarchy between the distances gets more pronounced, converging toward a value below one. We have confirmed that this is a general behavior for several other domains of the parameter space. 

The analysis performed supports the conclusion that microstate geometries with an angular momentum that is at a finite fraction of the cc bound must have a difference in scale between their inter-center distances.

\subsection{A particular solution}

Here we give the explicit form of the harmonic functions characterizing a BPS scaling microstate geometry with four Gibbons-Hawking centers. The solution has been found following the recipe detailed in Section \ref{subsec:SystematicProcedure}, taking the initial parameters from the region that optimizes the value of the entropy parameter according to the results of the numerical analysis. 

The solution is determined by the following harmonic functions,
\begin{equation}
\begin{split}
& V \:=\: \frac{1}{r_0} \:+\: \frac{1}{r_1} \:-\: \frac{2}{r_2} \:+\: \frac{1}{r_3} \\
& K^1 \:=\:  -\frac{36}{r_0} \:+\: \frac{100}{r_1} \:+\:\frac{18}{r_2} \:-\: \frac{38}{r_3}  \\
& K^2 \:=\:  \frac{278}{r_0} \:-\: \frac{4997}{r_1} \:-\: \frac{1702}{r_2} \:+\: \frac{220}{r_3}  \\
& K^3 \:=\:  \frac{344}{r_0} \:+\: \frac{342}{r_1} \:-\: \frac{2154}{r_2} \:+\: \frac{1644}{r_3}  \\
& L^1 \:=\:  1 \:-\: \frac{95632}{r_0} \:+\: \frac{1708974}{r_1} \:+\: \frac{1833054}{r_2} \:-\: \frac{361680}{r_3}  \\
& L^2 \:=\: 1 \:+\: \frac{12384}{r_0} \:-\: \frac{34200}{r_1} \:-\: \frac{19386}{r_2} \:+\: \frac{62472}{r_3}  \\
& L^3 \:=\:  1 \:+\: \frac{10008}{r_0} \:+\: \frac{499700}{r_1} \:-\: \frac{15318}{r_2} \:+\: \frac{8360}{r_3}  \\
& M \:=\: 2990.5 \:-\: \frac{1721376}{r_0} \:-\: \frac{85448700}{r_1} \:+\: \frac{8248743}{r_2} \:-\: \frac{6871920}{r_3}.
\label{NewSol}
\end{split}
\end{equation}

\noindent
The bubble equations can be solved numerically for the location of the centers,
\begin{equation}
z_1 \:=\: 5.9600\ldots \times 10^{-1} \,, \qquad z_2 \:=\: 1.1367\ldots \times 10^{-3}\,, \qquad z_3 \:=\: 7.5586\ldots \times 10^{-6} \, .
\end{equation}

\noindent
Performing an asymptotic expansion of $Z_I$ and $\mu$ we can obtain the three electric charges and the angular momentum of the solution, which can be read from the $\mathcal{O}(r^{-1})$ coefficients \cite{Bena:2006kb, Berglund:2005vb}
\begin{equation}
    \begin{split}
        & Q_1 \:=\: 1993340\\
        & Q_2 \:=\: 29014\\
        & Q_3 \:=\: 229906\\
        & J \:=\: -87655680.\\
    \end{split}
\end{equation}

\noindent
For these values of asymptotic charges the entropy parameter is
\begin{equation}
\mathcal{H}\:=\: 0.42\ldots
\end{equation}

\noindent
While this value is not close to 1, we can definitely affirm that it is far from 0. Thus, this microstate geometry corresponds to a rotating black hole whose angular momentum is significantly below the cc bound.

\subsubsection{Scaling the solution}
\label{subsubsec:scaling}

In general, one might need to break the axisymmetry of the configuration to scale the solutions. However, as it was proposed in \cite{Bena:2006kb}, axisymmetry can be preserved in the scaling process by slightly modifying the values of one of the parameters in the $K^I$ functions. Here, we choose to dial the value of $k_2^1$ but any other dipole charges could have worked. At each step one can check the bubble equations  and the absence of ctc's. The scaling process is summed up in the following table,
\bigbreak
\begin{center}
\begin{tabular*}{0.668\textwidth}{|c||c|c|c|c|}
\hline Sol & $k_2^1$ & \(z_1\) & \(\frac{z_1}{z_2}\) & \(\frac{z_2}{z_3}\) \\
\hline  1 & 100.00046  & $5.0152\times10^{-3}$ & $524.33$  & $150.38$  \\
\hline  2 & 100.0004639  & $4.6445\times10^{-6}$ & $524.33$  & $150.38$\\
\hline  3 & 100.0004639036 & $1.9403\times10^{-8}$ & $524.33$  & $150.38$\\
\hline  4 & 100.000463903615 & $1.3199\times10^{-10}$ & $524.33$  & $150.38$\\
\hline  5 & 100.0004639036151 & $3.5190\times10^{-12}$ & $524.33$  & $150.38$\\
\hline
\end{tabular*}
\end{center}

As explained in \cite{Bena:2006kb}, in the scaling process the microstate geometry develops a ``throat" that resembles the near-horizon geometry of an extremal black hole to increasing accuracy. The depth of this throat gets larger and larger as the cluster of centers shrinks. Since a BPS black hole has an infinite throat, during the scaling process the bubbling solution becomes more and more similar to the exterior of the black hole solution. Therefore, we have found a specific example of an asymptotically flat, scaling microstate geometry with four Gibbons-Hawking centers that corresponds to a microstate of a BMPV black hole with $\mathcal{H} \:=\: 0.42$.

\section{A supertube with three Gibbons-Hawking centers}

As we already mentioned in Section \ref{subsec:microgeom}, BPS scaling solutions with one supertube and three Gibbons-Hawking centers can be generated from three-supertube solutions in Taub-NUT. These configurations are interesting because they correspond to smooth horizonless microstate geometries in the D1-D5-P frame. We follow the same approach as in previous section. First, we explain how such solutions can be systematically generated and we perform a numerical analysis of the dependence of the entropy parameter on the initial parameters. Second, we present explicit examples of solutions with and without scale differences between the four centers.

To obtain our solutions one only needs to apply two generalized spectral flows to the original system of three supertubes. As we have seen, spectral flow transformations are responsible for decreasing the entropy parameter, so one may hope to find solutions with low angular momentum even without imposing a hierarchy of scales in the inter-center distances. 

\subsection{Exploration of the parameter space}

\subsubsection{Systematic generation of solutions}
\label{subsec:SystematicProcedure2}

We start from solutions that do not have a hierarchy of scales:

\begin{equation}
\begin{split}
\frac{z_1 - z_2}{z_3} \:\approx\: 1 , \\
\frac{z_2 - z_3}{z_3} \:\approx\: 1.
\end{split}
\label{distanceratios2} 
\end{equation}

The technique to generate these configurations is very similar to the method of \cite{Heidmann:2017cxt}, reviewed in detail in Section \ref{subsec:SystematicProcedure}. The only difference is that one of the generalized spectral flows is not applied. In a nutshell, the starting point is a three-supertube configuration with a Taub-NUT base space satisfying \eqref{distanceratios2}, which is characterized by seven parameters $k_1$, $k_2$, $k_3$, $\frac{Q_2^{(1)}}{Q_1^{(3)}}$, $\frac{Q_3^{(2)}}{Q_1^{(2)}}$, $\frac{Q_2^{(3)}}{Q_3^{(1)}}$ and $q_0$. Then, we apply any two generalized spectral flows. The corresponding parameters, say $\gamma^J$ and $\gamma^K$, are fixed by imposing a particular value for the Gibbons-Hawking integer charges generated in the process $q_J$ and $q_K$ which are free as long as $\sum q_a=1$. Then, we apply three gauge transformations to cancel the constant terms of the $K^I$. Finally, we truncate the constant term of the harmonic function $V$ to obtain a base space asymptotic to $\mathbb{R}^4$ and we round to integers all the charges in the harmonic functions. By systematizing this procedure it is possible to scan vast classes of solutions, parameterized by $k_1,k_2,k_3,\frac{Q_2^{(1)}}{Q_1^{(3)}},\frac{Q_3^{(2)}}{Q_1^{(2)}},\frac{Q_2^{(3)}}{Q_3^{(1)}},q_0$ and $q_J$. 

The same procedure can also be applied to solutions with a hierarchy of scales. As we saw in previous section, increasing the scale difference has a significant impact on the entropy parameter of four-GH-center solutions, which can reach values of the order of $\mathcal{H} \sim 0.5$. It is natural to ask how large this parameter can be for solutions with three Gibbons-Hawking centers and one supertube.

\subsubsection{Main results of the analysis}
\label{subsubsec:NumAnalysis2}

The details of the numerical analysis are contained in Appendix \ref{sec:numerics2}. After scanning relevant domains of the space of parameters, we have reached the following conclusions when looking for the best value of $\mathcal{H}$:

\begin{itemize}
\item The optimal location of the supertube is the outermost one: $(0,0,z_1)$. 
\item The Gibbons-Hawking charges $q_a$ should have the smallest possible absolute value, $|q_2|=|q_3|=|q_0|=1$, in agreement with what we found in Section \ref{subsubsec:NumAnalysis}.
\item All sign configurations for the initial dipole charges $k_a$ appear to be equally favored. We find that, when $k_2$ is taken negative, the entropy parameter reaches a maximum for a particular value of $\frac{k_1}{k_3}$, regardless of the values of the other parameters.
\item For aspect ratios satisfying \eqref{distanceratios2}, the maximal value of $\mathcal{H}$ is around $0.25$.
\end{itemize}

The analysis confirms what we anticipated: When only two generalized spectral flows are performed, the resulting solutions have lower angular momentum. Thus, one can reach a finite value of $\mathcal{H}$ even without a hierarchy of scales.

Of course, we just found in the previous section that hierarchic configurations can improve the value of the entropy parameter, at least for four GH centers. So we would like to investigate how adding a hierarchy of scales affects the angular momentum of solutions with one supertube. For that purpose, let us define the variable $m$ as we did in the previous section,
\begin{equation}
\begin{split}
\frac{z_1}{z_2} &\:\approx\: 10^m \\
\frac{z_2}{z_3} &\:\approx\: 10^m.
\end{split}
\end{equation}

\begin{figure}[h]
\centering
\includegraphics[width=100mm]{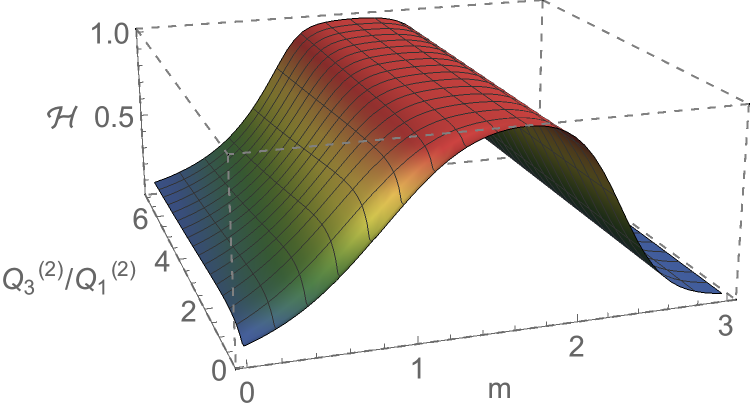}
\caption{Representation of the entropy parameter $\mathcal{H}$ as a function of the charge ratio $\frac{Q_3^{(2)}}{Q_1^{(2)}}$ and the order of magnitude of the inter-center distance ratio. The other parameters are $q_0 =q_2= 1$, $k_1= -k_2= k_3 = 1$, $\frac{Q_2^{(1)}}{Q_1^{(3)}} = 0.85$ and $\frac{Q_2^{(3)}}{Q_3^{(1)}} = 0.009$.}
\label{graphs_distanceratio2}
\end{figure}

We can then evaluate the value of the entropy parameter for a large set of solutions with different values of $m$ and the charge ratio $\frac{Q_3^{(2)}}{Q_1^{(2)}}$. The other parameters are fixed to optimal values according to the analysis performed for $m\approx0$. The result is very surprising. As the value of $m$ increases the value of the entropy parameter improves significantly and can stay arbitrarily close to $\mathcal{H}=1$ in a large region of the moduli space. This maximal value is obtained for $m\sim 1.5$, so the hierarchy of scales is not too pronounced. Unexpectedly, the value of the entropy parameter decreases if we go beyond that optimal hierarchy, see Fig.\ref{graphs_distanceratio2}.\\

Solutions with $m\sim 1.5$ are non-spinning. Indeed, one can find ctc-free scaling solutions with one supertube and three Gibbons-Hawking centers for which the spectral flow transformations completely annihilate the original angular momentum. However, those solutions typically have irrational charges. 

To obtain solutions with integer charges and fluxes, one has to first round these charges to nearby rational ones, and this typically brings back some angular momentum. However, the value of this angular momentum is proportional to the rounding, and hence can be made arbitrarily small by tightening the rounding. Hence, one can find regular scaling solutions with an entropy parameter infinitesimally close to one. In Section \ref{subsec:solnonspinning} we give an explicit example of such solutions.

\subsection{An example of a solution without scale differences}

Following the procedure outlined in Section \ref{subsec:SystematicProcedure2}  we can easily construct solutions with one supertube and three GH centers. For example:
\begin{equation}
\begin{split}
& V \:=\: \frac{1}{r_0} \:+\: \frac{1}{r_2} \:-\: \frac{1}{r_3} \\
& K^1 \:=\:  -\frac{184}{r_0} \:-\: \frac{60}{r_1} \:+\: \frac{27}{r_2} \:+\: \frac{361}{r_3}  \\
& K^2 \:=\:  -\frac{145}{r_0} \:+\: \frac{10909}{r_2} \:+\: \frac{5308}{r_3}  \\
& K^3 \:=\: \frac{1}{r_0} \:-\: \frac{68}{r_2} \:+\: \frac{67}{r_3} \\
& L^1 \:=\:  1 \:+\: \frac{145}{r_0} \:+\: \frac{741812}{r_2} \:+\: \frac{355636}{r_3}  \\
& L^2 \:=\: 1 \:+\: \frac{184}{r_0} \:-\: \frac{1300}{r_1} \:+\: \frac{1836}{r_2} \:+\: \frac{24187}{r_3}  \\
& L^3 \:=\:  1 \:-\: \frac{26680}{r_0} \:+\: \frac{2194116}{r_1} \:-\: \frac{294543}{r_2} \:+\: \frac{1916188}{r_3}  \\
& M \:=\: -8108 \:+\: \frac{13340}{r_0} \:+\: \frac{23769590}{r_1} \:-\: \frac{10014462}{r_2} \:+\: \frac{64192298}{r_3}.
\label{NewSol2}
\end{split}
\end{equation} 

\noindent
where $r_a$ are Euclidean three-dimensional distances measured from the centers at $(0,0,z_a)$. These locations are obtained solving numerically the bubble equations, which yield

\begin{equation}
z_1 \:=\: 1.0635\ldots \times 10^{-2} \,, \qquad z_2 \:=\: 7.1863\ldots \times 10^{-3}\,, \qquad z_3 \:=\: 3.5109\ldots \times 10^{-3}.
\end{equation}
The three global electric charges and the angular momentum are
\begin{equation}
    \begin{split}
        & Q_1 \:=\: 1097593\\
        & Q_2 \:=\: 24907\\
        & Q_3 \:=\: 6103449\\
        & J \:=\: 357140114.\\
    \end{split}
\end{equation}

\noindent
The entropy parameter of this solution is
\begin{equation}
\mathcal{H}\:\approx\: 0.24,
\end{equation}
which means that the angular momentum, $J$, is at \(87\%\) of its maximal value for those electric charges.

\subsubsection{Scaling solutions}

Following the procedure outlined in Section \ref{subsubsec:scaling}, we scale the solution by fine-tuning the value of $k_2^1$ . At each step in the scaling process, we solve the bubble equations and check for the absence of ctc's. The results are summed up in the following table:
\bigbreak
\begin{center}
\begin{tabular*}{0.700\textwidth}{|c||c|c|c|c|}
\hline Sol & $k_2^1$ & \(z_1\) & \(\frac{z_1-z_2}{z_3}\) & \(\frac{z_2-z_3}{z_3}\) \\
\hline  1 & -184.00003  & $1.2834\times10^{-3}$ & $0.98237$  & $1.0469$  \\
\hline  2 & -184.000034  & $3.6513\times10^{-5}$ & $0.98236$  & $1.0469$\\
\hline  3 & -184.00003411 & $2.2225\times10^{-6}$ & $0.98236$  & $1.0469$\\
\hline  4 & -184.000034117 & $4.0366\times10^{-8}$ & $0.98236$  & $1.0469$\\
\hline  5 & -184.000034117128 & $4.6524\times10^{-10}$ & $0.98236$  & $1.0469$\\
\hline  6 & -184.00003411712949 & $7.6773\times10^{-13}$ & $0.98236$ & $1.0469$\\
\hline
\end{tabular*}
\end{center}
\bigbreak
\subsection{A solution with very small angular momentum.}
\label{subsec:solnonspinning}

Here we build a solution with one supertube and three Gibbons-Hawking centers which has an entropy parameter $\mathcal{H} \sim 1$. For this purpose, we choose appropriately the scale difference between the inter-center distances and the values of the initial charges and dipole charges of the three supertubes to maximize the entropy parameter. Our procedure allows us to fine-tune the parameters to have $\mathcal{H}$ infinitesimally close to 1, and we present an example with $\mathcal{H}=0.999997$:
\begin{equation}
\begin{split}
& V \:=\: \frac{1}{r_0} \:+\: \frac{1}{r_2} \:-\: \frac{1}{r_3} \\
& K^1 \:=\:  -\frac{114}{r_0} \:-\: \frac{5}{r_1} \:-\: \frac{110}{r_2} \:+\: \frac{115}{r_3}  \\
& K^2 \:=\:  -\frac{111}{r_0} \:+\: \frac{4698}{r_2} \:+\: \frac{642}{r_3}  \\
& K^3 \:=\: \frac{3}{r_0} \:-\: \frac{87}{r_2} \:+\: \frac{84}{r_3} \\
& L^1 \:=\:  1 \:+\: \frac{333}{r_0} \:+\: \frac{408726}{r_2} \:+\: \frac{53928}{r_3}  \\
& L^2 \:=\: 1 \:+\: \frac{342}{r_0} \:+\: \frac{10}{r_1} \:-\:\frac{9570}{r_2} \:+\: \frac{9660}{r_3}  \\
& L^3 \:=\:  1 \:-\: \frac{12654}{r_0} \:+\: \frac{381142}{r_1} \:+\: \frac{516780}{r_2} \:+\: \frac{73830}{r_3}  \\
& M \:=\: -2557.5 \:+\:\frac{18981}{r_0} \:-\: \frac{381142}{r_1} \:+\: \frac{22479930}{r_2} \:+\: \frac{3100860}{r_3}.
\label{NewSol3}
\end{split}
\end{equation} 
The bubble equations give the positions of the centers:
\begin{equation}
z_1 \:=\: 7.3189\ldots \times 10^{-2} \,, \qquad z_2 \:=\: 3.6046\ldots \times 10^{-3}\,, \qquad z_3 \:=\: 9.7241\ldots \times 10^{-5}.
\end{equation}
The three charges and the angular momentum are:
\begin{equation}
    \begin{split}
        & Q_1 \:=\: 462987\\
        & Q_2 \:=\: 442\\
        & Q_3 \:=\: 362992\\
        & J \:=\: -16021,\\
    \end{split}
\end{equation}
giving, as advertised, an entropy parameter
\begin{equation}
\mathcal{H}\:=\: 0.999997\ldots.
\end{equation}
Thus, the angular momentum is at \(0.17\%\) of the cc bound.

\subsubsection{Scaling solutions}

We scale the solution by fine-tuning the value of $k_2^1$. At each step, we solve the bubble equations and check the absence of closed timelike curves. The scaling process is summed up in the following table:
\bigbreak
\begin{center}
\begin{tabular*}{0.6755\textwidth}{|c||c|c|c|c|}
\hline Sol & $k_2^1$ & \(z_1\) & \(\frac{z_1}{z_2}\) & \(\frac{z_2}{z_3}\) \\
\hline  1 & -113.999996  & $3.0729\times10^{-3}$ & $20.304$  & $37.068$  \\
\hline  2 & -113.99999583  & $9.2980\times10^{-5}$ & $20.304$  & $37.068$\\
\hline  3 & -113.999995825 & $5.3346\times10^{-6}$ & $20.304$  & $37.068$\\
\hline  4 & -113.9999958247 & $7.5857\times10^{-8}$ & $20.304$  & $37.068$\\
\hline  5 & -113.9999958246957 & $4.8195\times10^{-10}$ & $20.304$  & $37.068$\\
\hline
\end{tabular*}
\end{center}
\bigbreak


\section*{Acknowledgments}

We are indebted to Tom\'as Ort\'in, David Turton and Nick Warner for interesting discussions. The work of IB and PH is supported by the ANR grant Black-dS-String. The work of PH is supported by an ENS Lyon grant. The work of PFR was supported by the Severo Ochoa pre-doctoral grant SVP-2013-067903. This work has been supported by the Spanish Goverment grant FPA2015-66793-P (MINECO/FEDER, UE) and the Centro de Excelencia Severo Ochoa Program grant SEV-2016-0597. 


\appendix

\section{Solving the BPS equations}
\label{sec:BPSsolutions}

The action of the STU model of $\mathcal{N}=1$, $d=5$ supergravity is completely determined by the constant symmetric tensor $C_{IJK}=\vert \varepsilon_{IJK} \vert$. All the timelike-supersymmetric-field configurations of this theory have a conformastationary metric \cite{Gutowski:2004yv}
\begin{equation}
ds^2=-\left( Z_1 Z_2 Z_3 \right)^{-2/3} \left( dt+ k \right)^2 + \left( Z_1 Z_2 Z_3 \right)^{1/3} h_{mn} dx^m dx^n \, ,
\end{equation}

\noindent
where $h_{mn} dx^m dx^n$ is the metric of a hyper-K\"ahler manifold, while $Z_I$ and $k$ are respectively three functions and a 1-form taking values in this four-dimensional space. The remaining bosonic content consists of three vector fields satisfying
\begin{equation}
\label{eq:vectors}
A^I=-\frac{1}{Z_I} \left(dt+k\right) + B^I \, ,
\end{equation}

\noindent 
 and two scalars that can be conveniently parametrized as
 \begin{equation}
 e^{2\phi}=e^{2\phi_\infty} \frac{Z_0}{Z_1} \, , \qquad e^{2k}=e^{2k_\infty} \left( \frac{Z_3^2}{Z_1 Z_2} \right)^{1/2} \, ,
 \end{equation}
 \noindent
where $B^I$ is a 1-form in the hyper-K\"ahler space. These field configurations become solutions when the following set of \emph{BPS equations}, defined on the four-dimensional manifold, is satisfied 
\begin{eqnarray}
dB^I &=& \star_{(4)} dB^I \, , \\
\nabla^2_{(4)} Z_I &=& C_{IJK} \star_{(4)} \left( dB^J \wedge dB^K \right) \, , \\
dk+\star_{(4)} dk &=& Z_I dB^I \, ,
\end{eqnarray}

Therefore, the requirement that the solution is supersymmetric drastically simplifies the equations of motion of the theory to a linear system of PDE's on a manifold with Euclidean signature. Still, for general hyper-K\"ahler spaces this problem is a hard nut to crack. This is why, in order to make further progress, one usually chooses a specific, yet very general family of hyper-K\"ahler manifolds admitting a triholomorphic isometry. These are Gibbons-Hawking spaces \cite{Gibbons:1979zt}, whose metric is given by
\begin{equation}
h_{mn} dx^m dx^n = V^{-1} \left( d\psi+\chi \right)^2 + V \left( dx^2 + dy^2 + dz^2 \right) \, , \qquad \star_{(3)} dV=d \chi \, .
\end{equation}

\noindent
The integrability condition of the equation above implies that $V$ is harmonic in $\mathbb{R}^3$. We can make further progress if we assume that all matter fields are also independent of the isometric coordinate $\psi$. Then the functions and forms that characterize the solution can be further decomposed,
\begin{eqnarray}
B^I &=& -V^{-1} K^I (d\psi+\chi) + \breve{A}^I \, , \\
k &=& \mu (d\psi + \chi ) + \omega \, .
\end{eqnarray}

\noindent
Upon substitution in the system of BPS equations we find a set of differential equations for the three-dimensional \emph{seeds}
\begin{eqnarray}
\star_{(3)} dK^I &=& d \breve{A}^I \, , \\
\label{eq:omega}
\star_{(3)} d\omega &=& V dM - M dV + \frac{1}{2} \left(K^I d L_I - L_I d K^I \right)= <\Gamma , d \Gamma > \, , 
\end{eqnarray}

\noindent
and the following algebraic expressions for the building blocks that make up the solution,
\begin{eqnarray}
\mu &=& M + \frac{1}{2} V^{-1} L_I K^I + \frac{1}{6} V^{-2}C_{IJK} K^I K^J K^K \, , \\
\label{eq:ZI}
Z_I &=& L_I + \frac{1}{2} V^{-1} C_{IJK} K^J K^K \, .
\end{eqnarray}

\noindent
where $L_I$ and $M$ are harmonic functions in $\mathbb{R}^3$, $\nabla^2_{(3)} L_I=\nabla^2_{(3)} M=0$. Therefore, supersymmetric solutions admitting a spacelike isometry are completely specified in terms of 8 harmonic functions, $\Gamma = (V, K^I, L_I, M)$. Notice that the integrability condition of equation \eqref{eq:omega} yields the bubble equations

\begin{equation}
\label{eq:bubbleapp}
\sum_b \frac{ \langle \Gamma_a , \Gamma_b \rangle}{r_{ab}} = \langle\Gamma_\infty , \Gamma_a \rangle \, .
\end{equation}

It is convenient to define the \emph{quartic invariant} $\mathcal{I}_4$ which must satisfy the following inequality to avoid the presence of closed timelike curves \cite{Berglund:2005vb, Bena:2005va}

\begin{equation}
\mathcal{I}_4 \equiv Z_1 Z_2 Z_3 V - \mu^2 V^2 > 0\, .
\end{equation}

\noindent
This condition can be understood from the fact that the metric can be written as

\begin{equation}
\label{eq:metricexp}
ds^2= -f^2 dt^2 -2f^2 dt k +\frac{\mathcal{I}_4}{f^{-2} V^2} \left( d\psi + \chi - \frac{\mu V^2}{\mathcal{I}_4} \omega \right)^2 + f^{-1} V \left( d\vec{x} \cdot d\vec{x}-\frac{\omega^2}{\mathcal{I}_4} \right) \, ,
\end{equation}

\noindent
where we write $f^{-3} \equiv Z_1Z_2Z_3$.


\section{Numerical analysis of the entropy parameter of four-GH-center solutions}
\label{sec:numerics}
The aspect ratios of the solutions are fixed to:
\begin{equation}
\begin{split}
\frac{z_1}{z_2} &\:\approx\: 10^2 \\
\frac{z_2 }{z_3} &\:\approx\: 10^2.
\end{split}
\end{equation}
By generating such solutions using numerics, we want to describe the evolution of the entropy parameter $\mathcal{H}$ as a function of the nine degrees of freedom of the solutions $k_1$, $k_2$, $k_3$, $\frac{Q_2^{(1)}}{Q_1^{(3)}}$, $\frac{Q_3^{(2)}}{Q_1^{(2)}}$, $\frac{Q_2^{(3)}}{Q_3^{(1)}}$, $q_0$, $q_1$ and $q_2$. We decompose our analysis in three parts. We first analyze the entropy parameter by varying the initial supertube charges $\frac{Q_2^{(1)}}{Q_1^{(3)}}$, $\frac{Q_3^{(2)}}{Q_1^{(2)}}$ and $\frac{Q_2^{(3)}}{Q_3^{(1)}}$, with all the other parameters fixed. Then, we analyze the entropy parameter when varying $q_0$, $q_1$ and $q_2$. Finally, we analyze the entropy parameter as we vary the three initial dipole charges $k_1$, $k_2$ and $k_3$. \\
Each of the graphs is made by generating 2500 solutions following the procedure detailed in Section \ref{subsec:SystematicProcedure}. Because a configuration of parameters $k_1$, $k_2$, $k_3$, $\frac{Q_2^{(1)}}{Q_1^{(3)}}$, $\frac{Q_3^{(2)}}{Q_1^{(2)}}$, $\frac{Q_2^{(3)}}{Q_3^{(1)}}$, $q_0$, $q_1$ and $q_2$ can give different four-GH-center solutions, we take the final solution with the highest entropy parameter. Moreover, for readability reason, we smooth all the discrete graphs we initially obtained to have at the end a continuous curve.
\begin{itemize}
    \item The graphs in Fig.\ref{Numchargeratio} show the variations of the entropy parameter with the three ratios of supertube charges. The other parameters have been fixed to
    \begin{equation}
    \begin{split}
    & k_1 \:=\: - k_2 \:=\: k_3 \:=\: 1, \\
    & q_0 \:=\: q_1 \:=\: q_2 \:=\: 1.
    \end{split}
    \label{choicenumcharges}
    \end{equation}
    The entropy parameters can be greater than 15\% in many domains of charge ratios and more than 25\% in some small others.
     \begin{figure}
 \noindent\hfil\rule{0.7\textwidth}{.6pt}\hfil \\
 \centering
\begin{tabular}{ccc}
\addlinespace[1ex]
\subf{\includegraphics[width=70mm]{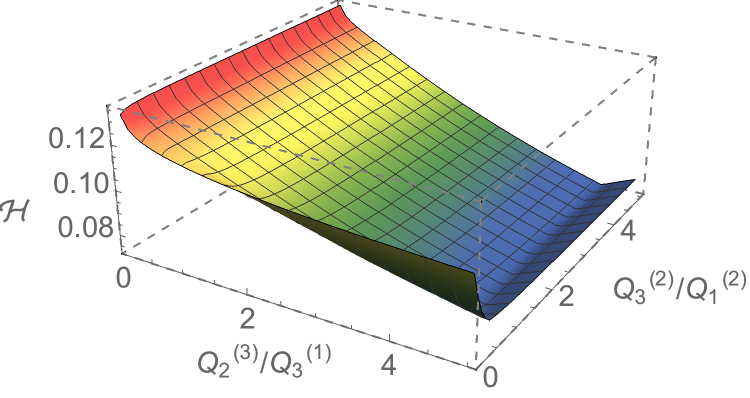}}
     {\\
     (a) $\frac{Q_2^{(1)}}{Q_1^{(3)}} \:=\: 0.1$}
&
\subf{\includegraphics[width=70mm]{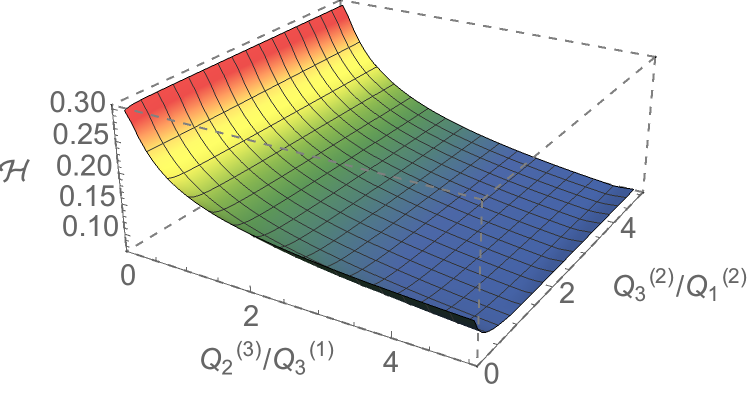}}
     {\\
     (b) $\frac{Q_2^{(1)}}{Q_1^{(3)}} \:=\: 0.5$}
\\
 \addlinespace[2ex]
\subf{\includegraphics[width=70mm]{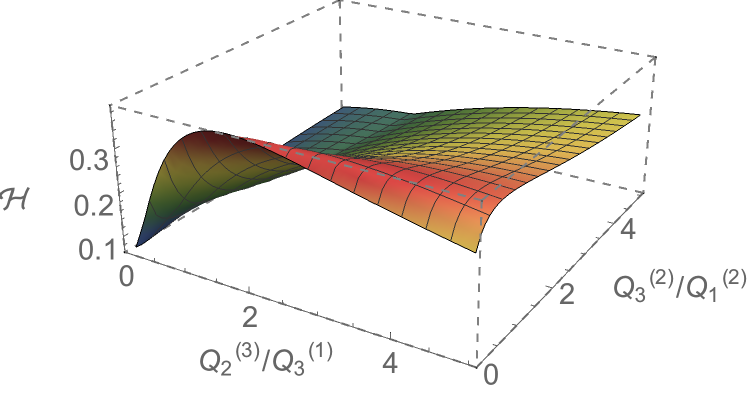}}
     {\\
     (c) $\frac{Q_2^{(1)}}{Q_1^{(3)}} \:=\: 1$}
&
\subf{\includegraphics[width=70mm]{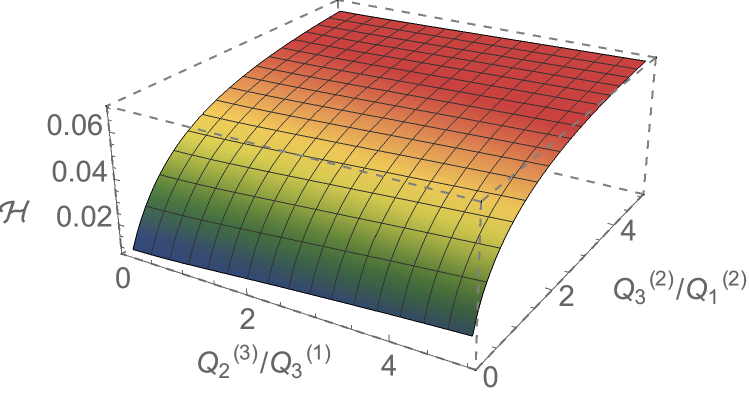}}
     {\\
     (e) $\frac{Q_2^{(1)}}{Q_1^{(3)}} \:=\: 5$}
\\
\addlinespace[1ex]
\end{tabular}
\noindent\hfil\rule{0.7\textwidth}{.6pt}\hfil
\caption{The entropy parameter $\mathcal{H}$ as a function of the charge ratios with $q$, $q_1$, $q_2$, $k_1$, $-k_2$ and $k_3$ equal to 1.}
\label{Numchargeratio}
\end{figure}
\item The graphs in Fig.\ref{NumVdep} illustrate the variation of the entropy parameter as a function of $q_0$, $q_1$ and $q_2$. We suppressed the values zero in the graphs. They correspond to three-GH-center and one-supertube solutions. The six other parameters have been fixed to
    \begin{equation}
    \begin{split}
    & k_1 \:=\: - k_2 \:=\: k_3 \:=\: 1, \\
    & 9 \frac{Q_2^{(1)}}{Q_1^{(3)}} \:=\: \frac{1}{2} \frac{Q_3^{(2)}}{Q_1^{(2)}} \:=\: \frac{Q_2^{(3)}}{Q_3^{(1)}} \:=\: 1.
    \end{split}
    \end{equation}
    However, we observed the same features for different values of charge ratios and dipole charges. The graphs show that for any value of $q_0$ the entropy is maximum when the absolute values of the charges are close to one. Furthermore the minimal Gibbons-Hawking charges (1,1,1 and -2) are the best choice to obtain four-GH-center solutions with low angular momentum. This is an unexpected feature. Indeed, in the five-center solution of \cite{Bena:2006kb}, the GH charges are close to each other and large. Our solutions do not share this feature.
 \begin{figure}
 \noindent\hfil\rule{0.7\textwidth}{.6pt}\hfil \\
 \centering
\begin{tabular}{ccc}
\addlinespace[1ex]
\subf{\includegraphics[width=70mm]{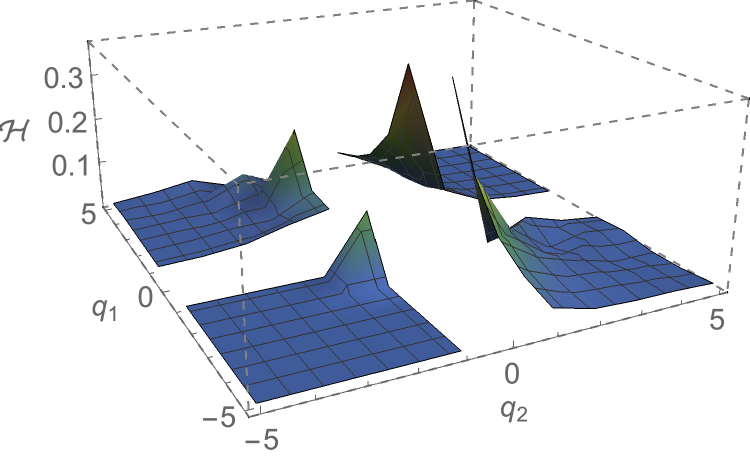}}
     {\\
     (a) $q_0 \:=\: 1$}
&
\subf{\includegraphics[width=70mm]{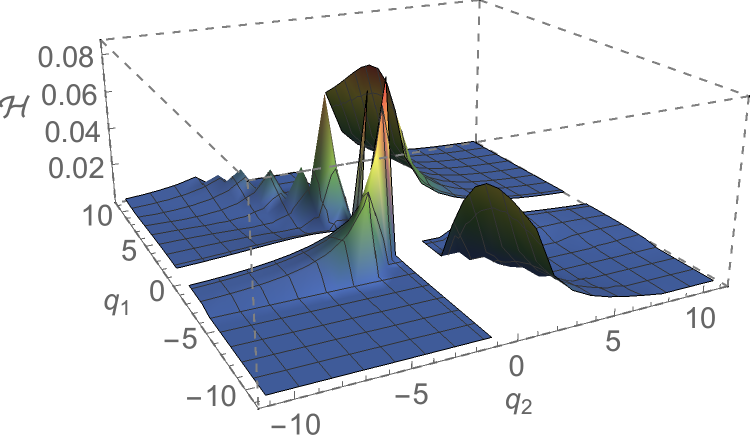}}
     {\\
     (b) $q_0 \:=\: 9$}
\\
 \addlinespace[2ex]
\subf{\includegraphics[width=70mm]{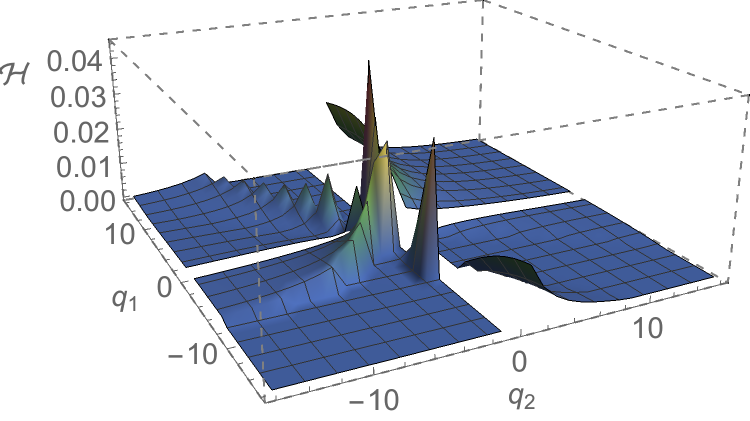}}
     {\\
     (c) $q_0\:=\: 13$}
&
\subf{\includegraphics[width=70mm]{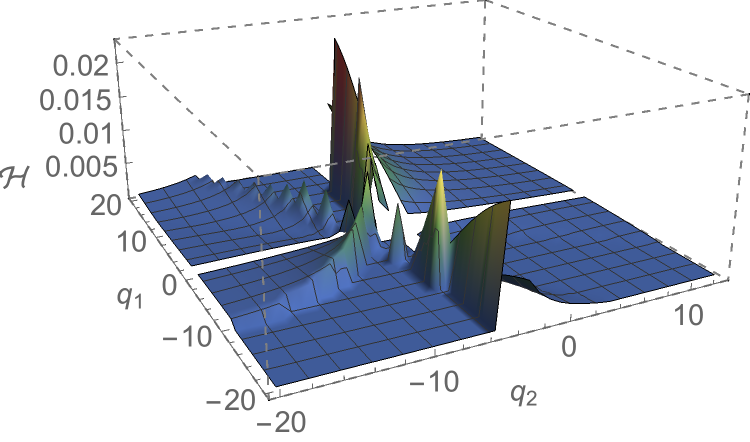}}
     {\\
     (d) $q_0 \:=\: 20$}
\\
\addlinespace[1ex]
\end{tabular}
\noindent\hfil\rule{0.7\textwidth}{.6pt}\hfil
\caption{The entropy parameter $\mathcal{H}$ as a function of the charges of V, $q_0$, $q_1$ and $q_2$ with $k_1$, $-k_2$, $k_3$ are equal to 1 and $\frac{Q_3^{(2)}}{Q_1^{(2)}}=2$, $\frac{Q_2^{(3)}}{Q_3^{(1)}}=1$ and $\frac{Q_2^{(1)}}{Q_1^{(3)}}=0.9$.}
\label{NumVdep}
\end{figure}
\item For the initial supertube dipole charges, we observed that the sign configuration given by \eqref{choicenumcharges} ($k_2$ negative, $k_1$ and $k_3$ positive) is the optimal one. With the two other sign configurations, we did not find domains of charges where the entropy parameter is above 0.1. For the rest of the analysis we focus on configurations with $k_2$ negative and $k_1$ and $k_3$ positive. By doing a quick analysis, we observed that the entropy parameter does not depend on the absolute value of $k_2$. The graphs in Fig.\ref{Numkdep} illustrate how the entropy parameter depends on the absolute value of the dipole charges $k_1$ and $k_3$. We vary also one charge ratio, $\frac{Q_2^{(1)}}{Q_1^{(3)}}$, keeping the other parameters fixed:
    \begin{equation}
    \begin{split}
    & q_0 \:=\: q_1 \:=\: q_2 \:=\: 1, \\
    & \frac{1}{2} \frac{Q_3^{(2)}}{Q_1^{(2)}} \:=\: \frac{Q_2^{(3)}}{Q_3^{(1)}} \:=\: 1.
    \end{split}
    \end{equation}
We remark that the entropy parameter depends essentially on the ratio $\frac{k_1}{k_3}$ and the entropy is maximum and far from 0 for one particular value of $\frac{k_1}{k_3}$. We observed the same kind of graph for different values of charge ratios. If one varies the value of $\frac{Q_2^{(1)}}{Q_1^{(3)}}$, the particular value of $\frac{k_1}{k_3}$ changes but the maximum value of the entropy parameter remains the same whereas if one varies the two other charge ratios both change. The maximum value of entropy parameter we observed is 0.3.
\begin{figure}
\centering
\begin{tabular}{ccc}
\addlinespace[1ex]
\subf{\includegraphics[width=70mm]{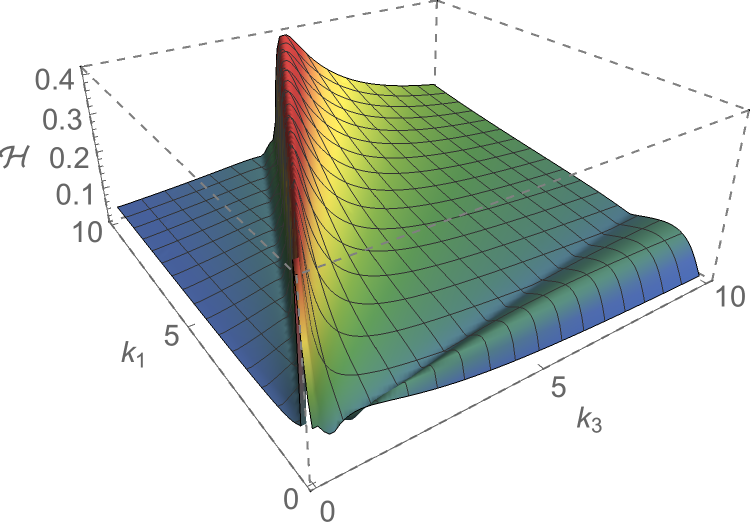}}
     {\\
     (a) $\frac{Q_2^{(1)}}{Q_1^{(3)}}=0.45$}
&
\subf{\includegraphics[width=70mm]{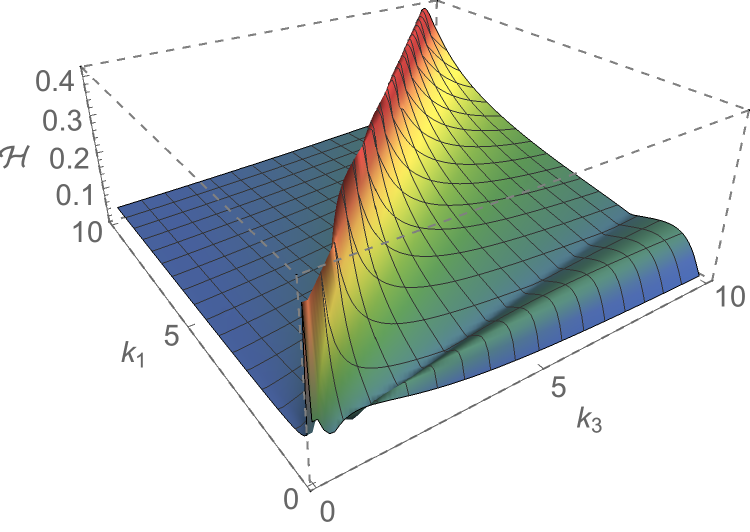}}
     {\\
     (b) $\frac{Q_2^{(1)}}{Q_1^{(3)}}=0.9$}
\\
 \addlinespace[2ex]
\subf{\includegraphics[width=70mm]{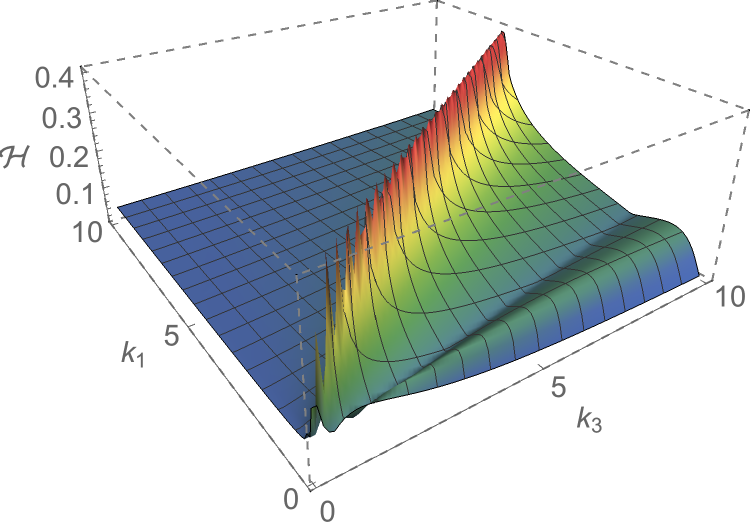}}
     {\\
     (c) $\frac{Q_2^{(1)}}{Q_1^{(3)}}=1.8$}
&
\subf{\includegraphics[width=70mm]{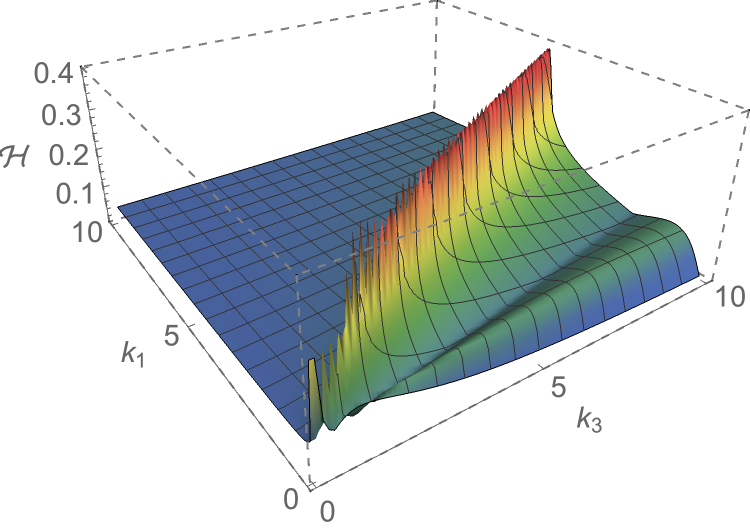}}
     {\\
     (d) $\frac{Q_2^{(1)}}{Q_1^{(3)}}=3.6$}
\\
\addlinespace[1ex]
\end{tabular}
\noindent\hfil\rule{0.7\textwidth}{.6pt}\hfil
\caption{The entropy parameter $\mathcal{H}$ as a function of the dipole charges $k_1$ and $k_3$ and one charge ratio $\frac{Q_2^{(1)}}{Q_1^{(3)}}$ with $q_0$, $q_1$, $q_2$ equal to 1 and $\frac{Q_3^{(2)}}{Q_1^{(2)}}=2$, $\frac{Q_2^{(3)}}{Q_3^{(1)}}=1$}
\label{Numkdep}
\end{figure}
\end{itemize}

To conclude, the numerical analysis shows that there exist large domains of supertube-charge ratios and supertube dipole charges where the entropy parameter of solutions satisfying \eqref{distanceratios3} is maximal and around 0.3. The only necessary conditions to have an angular momentum significantly below the cc bound is that the Gibbons-Hawking charges must be minimal and the dipole charge configuration of the generating three-supertube solution must be $k_1$ and $k_3$ positive and $k_2$ negative. Moreover, increasing the difference in scale between the inter-center distances does not affect how the entropy parameter varies with $k_1$, $k_2$, $k_3$, $\frac{Q_2^{(1)}}{Q_1^{(3)}}$, $\frac{Q_3^{(2)}}{Q_1^{(2)}}$, $\frac{Q_2^{(3)}}{Q_3^{(1)}}$, $q_0$, $q_1$ and $q_2$. It affects only the maximal value reachable as it was detailed in Section \ref{subsubsec:NumAnalysis}.

\section{Numerical analysis of the entropy parameter of solutions with one supertube and three Gibbons-Hawking centers}
\label{sec:numerics2}

We proceed the same way to analyse the entropy parameter of solutions with three Gibbons-Hawking centers and one supertube. We focus on solutions without scale differences between the inter-center distances:
\begin{equation}
\begin{split}
\frac{z_1-z_2}{z_3} &\:\approx\: 1 \\
\frac{z_2-z_3 }{z_3} &\:\approx\: 1.
\label{distanceratios3}
\end{split}
\end{equation}
According to the method used to generate them (see Section \ref{subsec:SystematicProcedure2}), the solutions depends on eight free parameters and the aspect ratios \eqref{distanceratios3}. We will also decompose our analysis in three parts. We first vary the initial supertube charges $\frac{Q_2^{(1)}}{Q_1^{(3)}}$, $\frac{Q_3^{(2)}}{Q_1^{(2)}}$ and $\frac{Q_2^{(3)}}{Q_3^{(1)}}$, with all the other parameters fixed. Then,  we analyze the entropy parameter as a function of $q_0$ and $q_J$, where $J$ is 1, 2 or 3 depending on which center is the supertube. Finally, we vary the three initial dipole charges $k_1$, $k_2$ and $k_3$. All the graphs have been generated as explained in the previous section.
\begin{itemize}
    \item First of all, we noticed that the localization of the supertube center compared to the three Gibbons-Hawking centers has a significant impact on the entropy parameter. The best configuration is when the supertube is not located between the Gibbons-Hawking centers. With our conventions, this means that the supertube center is the first center given by $(0,0,z_1)$. Indeed, we have found several domains of charges and dipole charges where the entropy parameter is above 0.15 for the three possible supertube locations. However, we have found that $\mathcal{H}$ has much higher values when the supertube is located at the first center.
    \item The graphs in Fig.\ref{Numanalchargeratio2SF} give the variations of the entropy parameter with the three initial charge ratios when the supertube is located at the first center. We have fixed the other parameters to be
    \begin{equation}
    \begin{split}
    & k_1 \:=\: - k_2 \:=\: k_3 \:=\: 1, \\
    & q_0 \:=\: - q_3 \:=\: 1.
    \end{split}
    \label{choicenumcharges2}
    \end{equation}
    We observe that when the initial charge ratio $\frac{Q_2^{(1)}}{Q_1^{(3)}}$ is between 0.4 and 1 and when $\frac{Q_2^{(3)}}{Q_3^{(1)}}$ is small, the entropy parameter can reach 0.25. This is the upper bound we found for a configuration which satisfies \eqref{choicenumcharges2} and \eqref{distanceratios3}.
    \begin{figure}
 \noindent\hfil\rule{0.7\textwidth}{.6pt}\hfil \\
 \centering
\begin{tabular}{ccc}
\addlinespace[1ex]
\subf{\includegraphics[width=70mm]{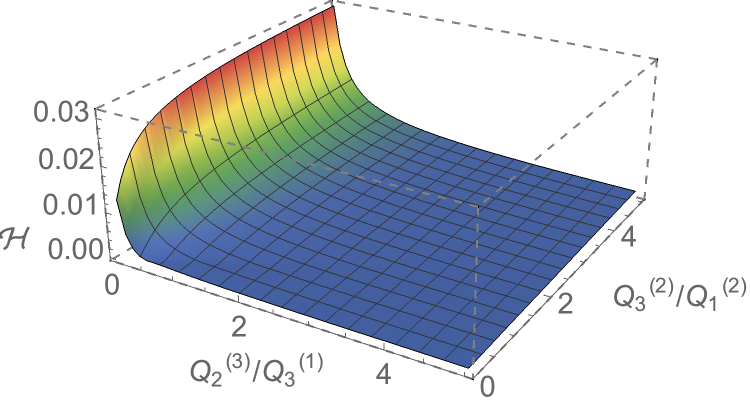}}
     {\\
     (a) $\frac{Q_2^{(1)}}{Q_1^{(3)}} \:=\: 0.1$}
&
\subf{\includegraphics[width=70mm]{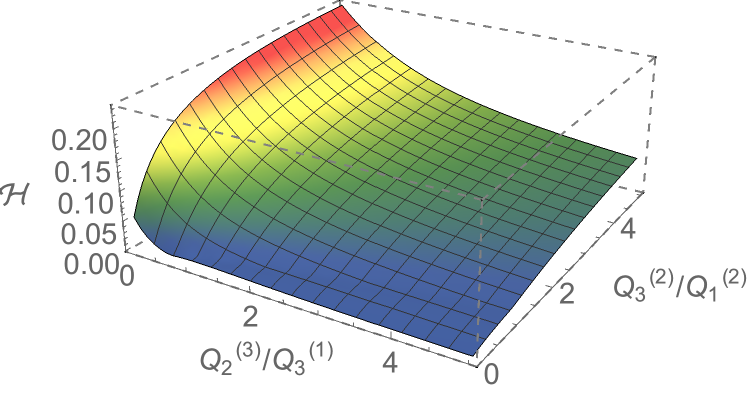}}
     {\\
     (b) $\frac{Q_2^{(1)}}{Q_1^{(3)}} \:=\: 0.5$}
\\
 \addlinespace[2ex]
\subf{\includegraphics[width=70mm]{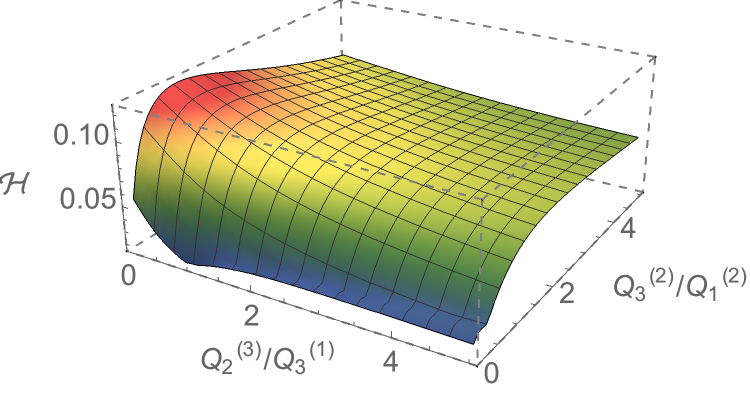}}
     {\\
     (c) $\frac{Q_2^{(1)}}{Q_1^{(3)}} \:=\: 1$}
&

\subf{\includegraphics[width=70mm]{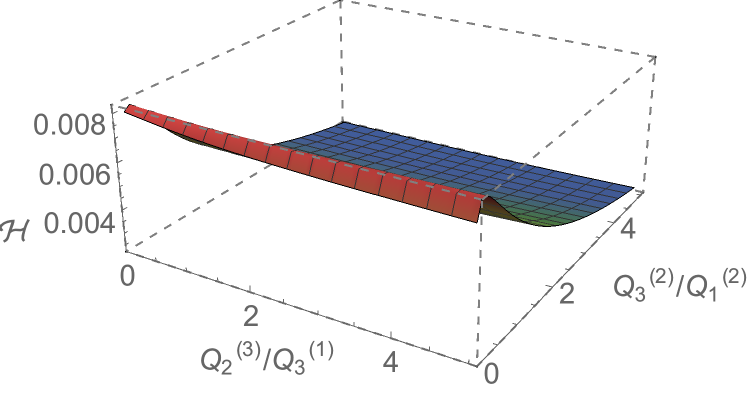}}
     {\\
     (e) $\frac{Q_2^{(1)}}{Q_1^{(3)}} \:=\: 5$}
     \\
\addlinespace[1ex]
\end{tabular}
\noindent\hfil\rule{0.7\textwidth}{.6pt}\hfil
\caption{The entropy parameter $\mathcal{H}$ as a function of the charge ratios with $q_0$, $q_2$, $k_1$, $-k_2$ and $k_3$ equal to 1.}
\label{Numanalchargeratio2SF}
\end{figure}
    \item Regarding the variation of the entropy parameter as a function of $q_0$ and $q_3$ ($q_2$ is fixed to satisfy $\Sigma \, q_a = 1$), we have observed the same features as in solutions with four Gibbons-Hawking centers: the higher the absolute value of the Gibbons-Hawking charges is, the lower is the entropy parameter. The graph in Fig.\ref{NumVdep2} shows the variation of the entropy parameter as a function of $q_0$ and $q_3$ for solutions satisfying \eqref{distanceratios3} and with
    \begin{equation}
    \begin{split}
    & k_1 \:=\: - k_2 \:=\: k_3 \:=\: 1, \\
    & 5\,\frac{Q_2^{(1)}}{Q_1^{(3)}}\:=\: 60\,\frac{Q_2^{(3)}}{Q_3^{(1)}} \:=\: \frac{1}{4} \frac{Q_3^{(2)}}{Q_1^{(2)}} \:=\: 1.
    \end{split}
    \end{equation}
\begin{figure}
\centering
\includegraphics[width=70mm]{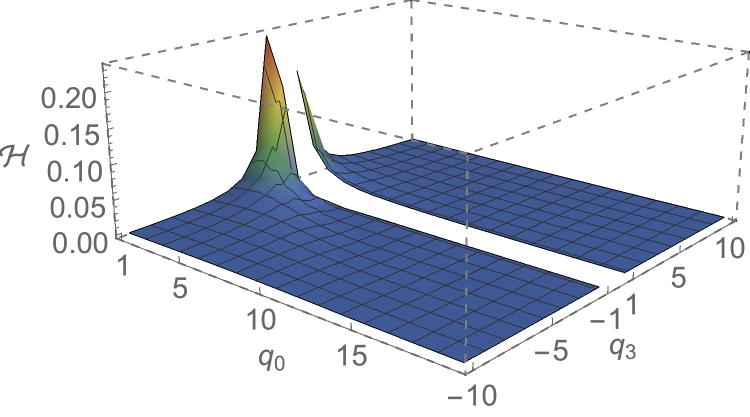}
\caption{The entropy parameter $\mathcal{H}$ as a function of the charges of V, $q_0$ and $q_3$ with $k_1$, $-k_2$, $k_3$ are equal to 1 and $\frac{Q_3^{(2)}}{Q_1^{(2)}}=4$, $\frac{Q_2^{(3)}}{Q_3^{(1)}}=0.06$ and $\frac{Q_2^{(1)}}{Q_1^{(3)}}=0.5$.}
\label{NumVdep2}
\end{figure}
We have observed similar variations for different initial charge ratios and dipole charges. Thus, $q_0 = 1$, $q_2 = 1$ and $q_3 = -1$ is the best configuration to optimize the entropy parameter.
\item Varying the initial supertube dipole charges, we have again observed exactly the same features as in solutions with four Gibbons-Hawking centers. The best sign configuration is when $k_2$ is negative and when $k_1$ and $k_3$ are positive. Moreover, the entropy parameter does not depend significantly on the absolute value of $k_2$ and it only depends on $\frac{k_1}{k_3}$. It also reaches a maximum for a particular value of the ratio $\frac{k_1}{k_3}$. The value and the location of the maximum depends on the values of the supertube charge ratios. The graphs in Fig.\ref{Numkdep2} illustrate these conclusions. We built solutions and computed their entropy as a function of the absolute value of the dipole charges $k_1$ and $k_3$ and one charge ratio $\frac{Q_2^{(1)}}{Q_1^{(3)}}$. The other parameters have been fixed to
    \begin{equation}
    \begin{split}
    & q_0 \:=\: q_1 \:=\: q_2 \:=\: 1. \\
    & \frac{1}{4} \frac{Q_3^{(2)}}{Q_1^{(2)}} \:=\: 60 \frac{Q_2^{(3)}}{Q_3^{(1)}} \:=\: 1.
    \end{split}
    \end{equation}
We have analyzed the entropy parameter for charge ratios different from the one above. The upper bound of all the maxima we observed is 0.25.
\begin{figure}
\centering
\begin{tabular}{ccc}
\addlinespace[1ex]
\subf{\includegraphics[width=70mm]{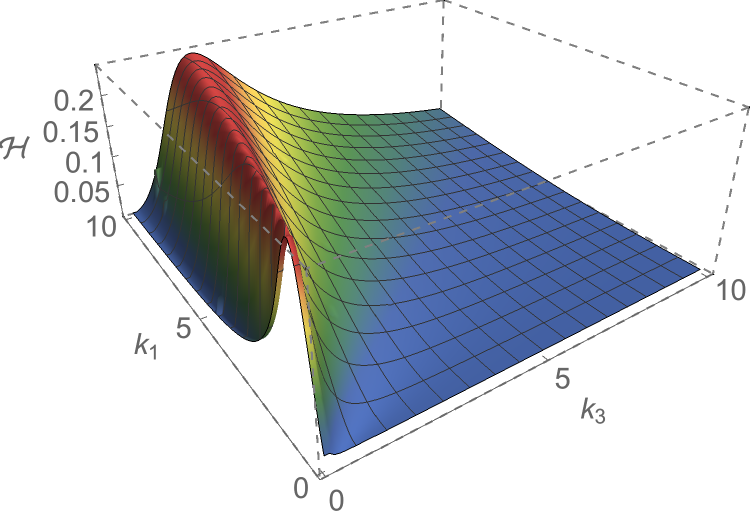}}
     {\\
     (a) $\frac{Q_2^{(1)}}{Q_1^{(3)}}=0.25$}
&
\subf{\includegraphics[width=70mm]{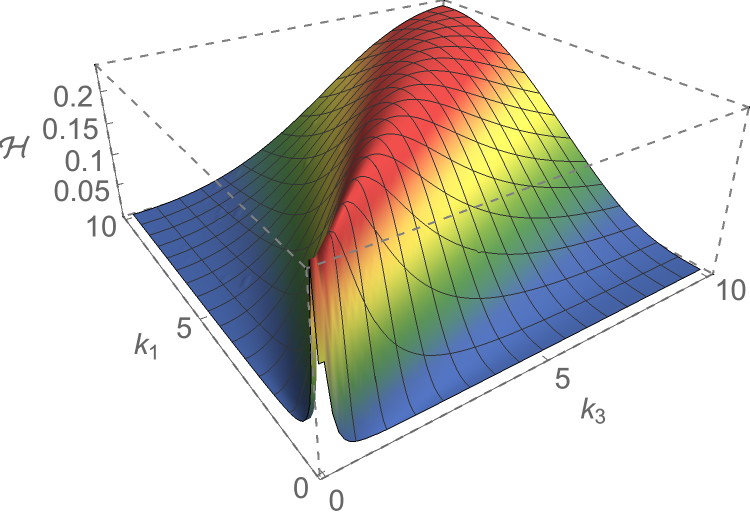}}
     {\\
     (b) $\frac{Q_2^{(1)}}{Q_1^{(3)}}=0.5$}
\\
 \addlinespace[2ex]
\subf{\includegraphics[width=70mm]{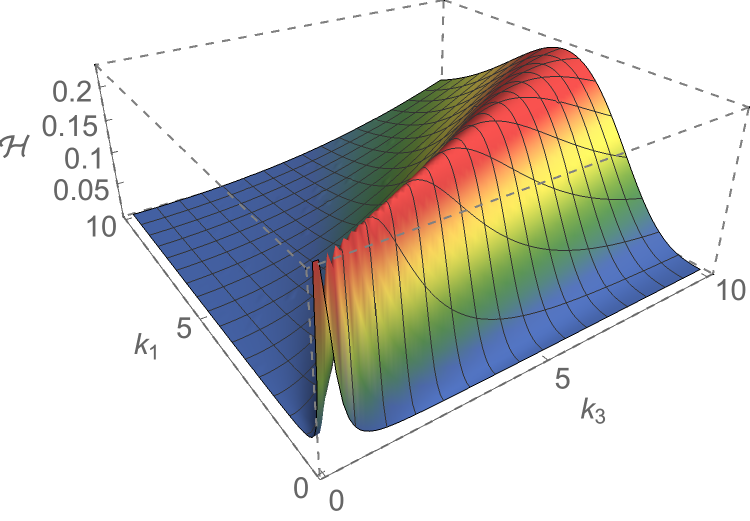}}
     {\\
     (c) $\frac{Q_2^{(1)}}{Q_1^{(3)}}=1$}
&
\subf{\includegraphics[width=70mm]{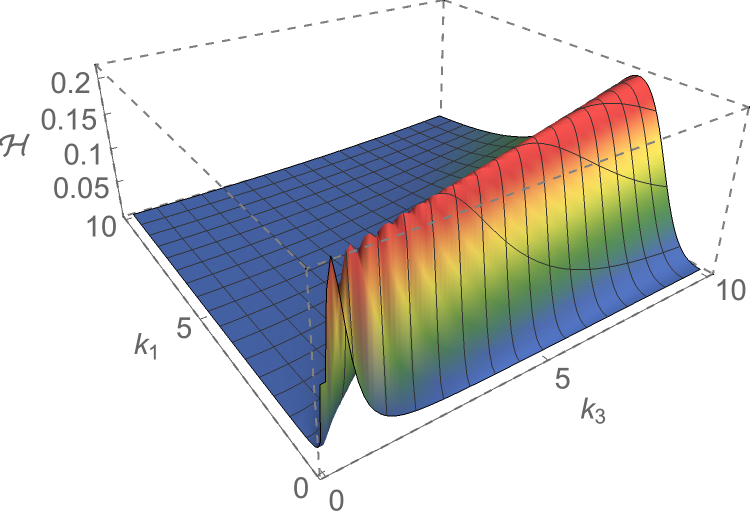}}
     {\\
     (d) $\frac{Q_2^{(1)}}{Q_1^{(3)}}=1.5$}
\\
\addlinespace[1ex]
\end{tabular}
\noindent\hfil\rule{0.7\textwidth}{.6pt}\hfil
\caption{The entropy parameter $\mathcal{H}$ as a function of the dipole charges $k_1$ and $k_3$ and one charge ratio $\frac{Q_2^{(1)}}{Q_1^{(3)}}$ with $q_0$, $q_1$, $q_2$ equal to 1 and $\frac{Q_3^{(2)}}{Q_1^{(2)}}=4$, $\frac{Q_2^{(3)}}{Q_3^{(1)}}=0.06$}
\label{Numkdep2}
\end{figure}
\end{itemize}

The numerical analysis shows that solutions with one supertube and three Gibbons-Hawking centers do not need to have a scale difference between the inter-center distances to have an entropy parameter above 0.1. If one chooses minimal Gibbons-Hawking charges and $k_2$ negative, $k_1$ and $k_3$ positive, one can find domains of parameters where the entropy is around 0.2.

\newpage

\bibliography{references}
\bibliographystyle{utphysmodb}

\end{document}